%% file: Main.tex
\def\single{0} 
\def\onappendix{1} 
\def\arxivdisclaimer{1} 

\if\single1
\documentclass[draftcls,onecolumn,12pt]{IEEEtran}
\else
\documentclass[journal]{IEEEtran}
\fi
\usepackage{authblk}
\usepackage{amsmath}
\usepackage{amssymb}
\usepackage{amsthm}
\usepackage{mathabx}
\usepackage{cite}
\usepackage{float}
\usepackage{todonotes}
\usepackage{multirow}
\usepackage{graphicx}
\usepackage{enumerate}
\usepackage{wrapfig}
\usepackage{setspace}
\usepackage{soul,color}
\usepackage{soul}

\soulregister\cite7
\soulregister\ref7
\soulregister\pageref7

\if\arxivdisclaimer0

\else
\usepackage{hyperref}
\fi

\usepackage{cleveref}
\usepackage{algorithm}
\usepackage[noend]{algpseudocode}
\usepackage{xpatch}
\usepackage{verbatim}
\usepackage{bbm}
\usepackage[normalem]{ulem} 
\usepackage{epstopdf}
\usepackage{subcaption}

\usepackage[user]{zref}
\newcommand{\sot}[1]{}

\newcounter{revc}
\makeatletter \zref@newprop{revcontent}{} \zref@addprop{main}{revcontent}
\zref@newprop{revsec}{} \zref@addprop{main}{revsec}
\zref@newprop{revpage}{} \zref@addprop{main}{revpage}

\if\single1
\newcommand{\revi}[2]{
\zref@setcurrent{revsec}{\thesection}%
\zref@setcurrent{revpage}{\thepage}%
\zref@setcurrent{revcontent}{#2}%
\refstepcounter{revc}%
\label{#1}%
\zlabel{#1}%
\uline{#2}%
}

\newcommand{\revinu}[2]{%
\zref@setcurrent{revsec}{\thesection}%
\zref@setcurrent{revcontent}{#2}%
\refstepcounter{revc}%
\zlabel{#1}%
\label{#1}
#2 }

\newcommand{\revr}[2]{%
\zref@setcurrent{revsec}{\thesection}%
\zref@setcurrent{revcontent}{#2}%
\refstepcounter{revc}%
\zlabel{#1}%
\label{#1} \sot{#2}} \makeatother
\newcommand{\revp}[1]{\zref[revcontent]{#1}}

\else

\newcommand{\revi}[2]{
\zref@setcurrent{revsec}{\thesection}%
\zref@setcurrent{revpage}{\thepage}%
\zref@setcurrent{revcontent}{#2}%
\refstepcounter{revc}%
\label{#1}%
\zlabel{#1}%
#2%
}

\newcommand{\revinu}[2]{%
\zref@setcurrent{revsec}{\thesection}%
\zref@setcurrent{revcontent}{#2}%
\refstepcounter{revc}%
\zlabel{#1}%
\label{#1}
#2 }

\newcommand{\revr}[2]{%
\zref@setcurrent{revsec}{\thesection}%
\zref@setcurrent{revcontent}{#2}%
\refstepcounter{revc}%
\zlabel{#1}%
\label{#1} \sot{#2}} \makeatother
\newcommand{\revp}[1]{\zref[revcontent]{#1}}

\fi
\makeatletter

\usepackage{tikz}
\usepackage{pgfplots}
\usepackage{pgfplotstable}
\usepgfplotslibrary{colorbrewer}
\pgfplotsset{compat=1.8}

\tikzstyle{arrow} = [thick,->,>=stealth]

\pgfplotsset{compat=1.11,
    /pgfplots/ybar legend/.style={
    /pgfplots/legend image code/.code={%
       \draw[##1,/tikz/.cd,yshift=-0.25em]
        (0cm,0cm) rectangle (3pt,0.8em);},
   },
}

\pgfplotsset{vasymptote/.style={
    before end axis/.append code={
        \draw[solid] ({rel axis cs:0,0} -| {axis cs:#1,0})
        -- ({rel axis cs:0,1} -| {axis cs:#1,0});
    }
}}

\usepackage[acronym]{glossaries}
\glsdisablehyper

\newacronym{2D}{2D}{two-dimensional}
\newacronym{3GPP}{3GPP}{3rd generation partnership project}
\newacronym{5G}{5G}{5th generation}
\newacronym{6G}{6G}{6th generation}
\newacronym{AoA}{AoA}{angle of arrival}
\newacronym{AoD}{AoD}{angle of departure}
\newacronym{AAL}{AAL}{array aperture line}
\newacronym{AR}{SARA DFT}{SARA DFT-based reconstruction}
\newacronym{AWGN}{AWGN}{additive white Gaussian noise}
\newacronym{CDF}{CDF}{cumulative density function}
\newacronym{CCDF}{CCDF}{complementary cumulative density function}
\newacronym{CFAR}{CFAR}{constant false alarm rate}
\newacronym{CRLB}{RCA}{root Cram\'er-Rao lower bound approximation}
\newacronym{DFT}{DFT}{discrete Fourier transform}
\newacronym{DL}{DL}{downlink}
\newacronym{UL}{UL}{uplink}
\newacronym{eMBB}{eMBB}{enhanced mobile broadband}
\newacronym{InF}{InF}{indoor factory}
\newacronym{iid}{i.i.d.}{independent and identically distributed}
\newacronym{JCAS}{ISAC}{integrated sensing and communications}
\newacronym{UMa}{UMa}{urban macro}
\newacronym{ML}{ML}{machine learning}
\newacronym{LAD}{NAF}{normalized angular frequency}
\newacronym{LR}{SARA Conv.}{SARA Convolution}
\newacronym{MIMO}{MIMO}{multiple input multiple output}
\newacronym{MDL}{MDL}{minimum description length}
\newacronym{MMSE}{MMSE}{minimum mean square error}
\newacronym{NN}{NN}{neural network}
\newacronym{RF}{RF}{radio frequency}
\newacronym{RMSE}{RMSE}{root mean square error}
\newacronym{SARA}{SARA}{sampling and reconstructing angular domain}
\newacronym{SINR}{SINR}{signal to interference plus noise ratio}
\newacronym{SNR}{SNR}{signal to noise ratio}
\newacronym{TTI}{TTI}{time transmission interval}
\newacronym{TP}{TP}{topographic prominence}
\newacronym{TSDCE}{TSDCE}{Transformed Spatial Domain Channel Estimation}
\newacronym{UCyA}{UCyA}{uniform cylindrical array}
\newacronym{ULA}{ULA}{uniform linear array}
\newacronym{URA}{URA}{uniform rectangular array}
\newacronym{URLLC}{URLLC}{ultra-reliable low-latency communications}
\newacronym{mmWave}{mm-Wave}{millimeter wave}

\newtheorem{remark}{Remark}
\newtheorem*{remark*}{Remark}

\newtheorem{lemma}{Lemma}
\newtheorem{theorem}{Theorem}

\begin{document}

\title{Sampling and Reconstructing Angular Domains with Uniform Arrays}


\author{Silvio Mandelli, Marcus Henninger and Jinfeng Du, \emph{Member, IEEE}
\thanks{Silvio Mandelli and Marcus Henninger are with
Nokia Bell Laboratories, 70469 Stuttgart, Germany, (e-mail:
silvio.mandelli@nokia-bell-labs.com, marcus.henninger@nokia.com).}
\thanks{Jinfeng Du is with Nokia Bell Laboratories, Murray Hill, NJ 07974, USA,
(e-mail: jinfeng.du@nokia-bell-labs.com).}
\if\arxivdisclaimer1
\thanks{This work has been submitted to the IEEE for possible publication. Copyright may be transferred without notice, after which this version may no longer be accessible.}
\fi
}

\maketitle

\input{Content/Abstract}
\input{Content/Introduction}
\input{Content/SamplingAngles}
\input{Content/AngularReconstruction}

\input{Content/ULAReconstruction}

\input{Content/Simulation}

\input{Content/MultipleTargets}
\input{Content/2DExample}

\input{Content/Conclusion}

\input{Content/Acknowledgements}
\if\onappendix1
\input{Content/Appendixes}
\else
\fi

\bibliographystyle{IEEEtran}
\bibliography{references}

\if\single1
\input{Content/RevisionAnswers_secondReview}
\fi

\end{document}

%% file: Content/Abstract.tex
\if\single1
\vspace{-2cm}
\fi
\begin{abstract}
The surge of massive antenna arrays in wireless networks calls for the adoption of analog/hybrid
array solutions, where multiple antenna elements are driven by a common radio front end to form a beam
along a specific angle in order to maximize the beamforming gain.
Many heuristics have been proposed to sample the angular domain by trading off between sampling step size and overhead, where arbitrarily small angular step size is only attainable with infinite sampling overhead.  
We show that, for uniform linear and rectangular arrays, lossless reconstruction of the array's angular responses at arbitrary angular precision is possible using a finite number of samples without resorting to assumptions of angular sparsity.  
The proposed method, \gls{SARA},
defines how many and which angles to be sampled and the corresponding reconstruction.
This general solution to scan the angular domain can therefore be applied not only to beam acquisition and channel estimation, but also to radio imaging techniques, making it a candidate for future integrated sensing and communications (ISAC).
Extensive simulation results for target detection and radio imaging have demonstrated clear advantages of SARA over other considered baselines, both in terms of angular reconstruction performance and computational complexity.
\end{abstract}
\if\single1
\vspace{-0.5cm}
\fi
\begin{IEEEkeywords}
Angular sampling, Angular interpolation, Analog/Hybrid beamforming, Radar imaging
\end{IEEEkeywords}
\glsresetall
\if\single1
\vspace{-0.5cm}
\fi

%% file: Content/Introduction.tex
\section{Introduction}\label{sec:Intro}

The next generation of wireless networks  will enable technological advancements such as human-less factories~\cite{harish20206g,liu2019wireless} and \gls{JCAS} where radio frequency imaging of passive objects~\cite{harish20206g} could be performed on top of the legacy communications operations.
The large antenna arrays deployed for
massive \gls{MIMO} communications can be leveraged for fine angular resolution to separate nearby targets as well as high beamforming gains to compensate the two-way backscatter propagation loss~\cite{wild2021joint}.
%
Since cost and power
consumption of such massive arrays and their associated
transceiver chains may become prohibitive, 
analog beamforming~\cite{wang2009beam,hur2013millimeter} and hybrid beamforming~\cite{zhang2005variable,alkhateeb2013hybrid} have been proposed to drive multiple antenna elements by each radio front end via a network of analog phase shifters. 
\subsection{Motivation}\label{subsec:Motivation}
Several beam training algorithms were proposed in the literature 
to design the beamforming coefficients to focus transmission or reception on an arbitrary incident angle to the antenna array~\cite{wang2009beam,hur2013millimeter,zhang2005variable,alkhateeb2013hybrid,arora2019hybrid,rajamaki2019analog,rajamaki2020hybrid,roger2021low,roger2021fast,zhang2021gridless,lin2021joint}.
Accordingly, a specific set of beamforming coefficients corresponds to sampling the angular domain of that array in a specific point.
%
Many heuristic approaches have been proposed to determine the minimum number and which angles must be scanned in order to probe the full angular domain with desired accuracy for channel estimation or angular estimation purposes.
%
For example, the concept of \gls{DFT} beamspace has been introduced in~\cite{zoltowski1996closed} to represent the angular capabilities of a \gls{ULA} and \gls{URA}, proposing a solution for the set of angles that must be sampled. This concept has been extended to analog/hybrid arrays recently for channel estimation purposes~\cite{roger2021fast,zhang2021gridless,roger2021low}. \revi{rev:TSDCE}{In particular, the algorithm proposed in~\cite{roger2021low} and~\cite{roger2021fast} performs channel estimation in the \gls{DFT} domain, but requires additional overhead due to estimating \gls{2D} transmit-receive angular pairs instead of a single angle, corresponding to the scenario of sensing acquisitions with an antenna array.} Finally, the prior analysis makes the common assumption of antenna arrays with half wavelength spacing, that does not allow generalization of their conclusions to either a generic \gls{ULA} or \gls{URA}. 

Differently from previous works on beamforming at \gls{mmWave} at higher frequencies, passive sensing applications in \gls{6G} networks require the full image of the sensed environment rather than a sparse angular representation, that is typically assumed~\cite{alkhateeb2013hybrid}.
This raises the challenge of interpolating the available angular acquisitions, determining the angular representation of the channel/environment at every possible angle, as if infinite angular acquisitions were possible.
The interpolation and reconstruction literature~\cite{lewitt1983reconstruction} has largely ignored the non-linearity of the generated phase shifts of \glspl{ULA} and \glspl{URA} at each antenna element with respect to the incident angle. For example, among the few which consider angular domains, the interpolation based on the effective aperture distribution function (EADF)~\cite{landmann2004efficient} proposes uniform sampling in the angular domain over the angular intervals of interest. However, in this work we show that this leads to distortion in the angular response reconstruction. 
\revi{rev:newRef2}{More recent work accounts for this non-linearity for hybrid \glspl{UCyA} 
in~\cite{lin2020nested}, where the authors proposed an efficient tensor-based angular estimation algorithm exploiting sparse array theory, achieving better resolution and lowering the number of required radio front ends.
Similarly, in~\cite{lin2021joint}, the authors considered linear angular interpolation for hybrid \gls{UCyA} angular array responses. However, both~\cite{lin2021joint} and~\cite{lin2020nested} focus on \gls{UCyA}, making the angular domain representation, its sampling, and its reconstruction very different from \gls{ULA} and \gls{URA}. In particular, the proposed linear interpolation cannot achieve lossless angular domain reconstruction, which is provided by the formulae derived in this work.}
Recent work on radio imaging with arrays~\cite{rajamaki2020hybrid} performs acquisitions with fine sampling of the angular domain, generating many (redundant) acquisitions with large overhead and reducing the amount of radio resources available for \gls{JCAS}. Moreover, these long acquisitions in dynamic scenarios would cause a signal change during the acquisitions, distorting the desired image, as will be discussed later. 

\revi{rev:ProblemSummary}{
Summarizing the problem, with the surge of analog and hybrid arrays in wireless communications, determining which angles to be sampled is critical to reduce acquisition time without losing the channel and environment information that can be gathered by the antenna array. Then, proper interpolation of the available angular samples is pivotal for imaging applications in future wireless networks.
}

\subsection{Our contributions}
In this work we leverage the Fourier duality between the space where \gls{ULA} antennas are placed and the \gls{LAD}, defined as the sine of the incident angle  with a linear coefficient depending on the antenna spacing and wavelength. This allows us to 
\begin{itemize}
    \item derive the \gls{SARA} criterion for \gls{ULA} and \gls{URA} angular response acquisitions, determining  how many and which angles must be sampled such that lossless reconstruction can be done for any possible angle without resorting to the channel sparsity assumption. \gls{SARA} is built by applying the Nyquist-Shannon sampling theorem~\cite{shannon1949communication} to the array space and the \gls{LAD} domains. Then, the Whittaker-Shannon interpolation has been used as starting point to derive the formulas to reconstruct the angular response. The final solution is shown to coincide with the trigonometric interpolation~\cite{jackson1913accuracy} of the angular samples, if they are taken uniformly in the \gls{LAD} domain;
    \item prove that a computationally efficient implementation of the lossless reconstruction can be obtained as a particular case of \gls{DFT} interpolation~\cite{jain1979high};
    \item compare via extensive numerical simulations the performance of \gls{SARA} in both angular estimation and RF imaging against interpolation baselines and the multiple signal classification (MUSIC) algorithm~\cite{schmidt1986multiple}, showing \gls{SARA} approaches the  \gls{CRLB} for \gls{LAD} estimation of an impulsive target~\cite{baronkin2001cramer}. 
\end{itemize}
{The proposed \gls{SARA} is particularly useful with imaging techniques, where one is not interested in a sparse representation of the channel, but rather in its overall response at every angle visible by the antenna array, like in~\cite{rajamaki2019analog,rajamaki2020hybrid}.}  
However, we show in our numerical evaluations that the proposed non-sparse reconstruction is valid also to estimate the sparse angular components of wireless channels, that is the typical assumption for \gls{mmWave} and higher carrier frequency systems~\cite{alkhateeb2013hybrid}.

\subsection{Paper organization}

The rest of the paper is organized as follows. 
Models and preliminaries are given in Section~\ref{sec:ArrayAngularDomains}. The \gls{SARA} angular sampling and reconstruction over a generic \gls{ULA} is formalized in Section~\ref{sec:AngularReconstruction}. 
Section \ref{sec:ULAReconstruction} elaborates on a few practical applications of \gls{SARA}: \gls{ULA} only at transmitter/receiver, \gls{ULA} at both ends, and generalization to \gls{URA} scenarios.
Section \ref{sec:Simulation} presents numerical experiments of single and multiple target angular estimation in a mono-static sensing scenario with full-duplex \gls{ULA} scans in the azimuth domain, as well as a \gls{2D} imaging example to visualize each step of \gls{SARA}.

\textbf{Notation:} row vectors are in bold lowercase and scalars are unbolded. The $n$-th
element of a vector $\mathbf{v}$ is denoted as $\mathbf{v}_n$. A row vector of $N$ zero (or unitary) elements is denoted as $\mathbf{0}_N$ (or $\mathbf{1}_N$). We define $\mathbf{I}_N$ as an identity matrix of $N$x$N$ elements.
The element-wise (Hadamard) product of two vectors is denoted by $\odot$. 
Sets are denoted in mathematical italic capital letters (e.g., $\mathcal{L}, \mathcal{D}$), 
while $\mathbb{Z}, \mathbb{N}_0$ are reserved for sets of integers, and natural numbers with zero, respectively. 
The rectangle function $\text{rect}(x)$ equals 1 for $x \in [-0.5, 0.5]$ and zero elsewhere. $\delta(x)$ is the Dirac delta function. The modulo function is defined as $\text{mod}_n(x) = x - n \lfloor x/n \rfloor$.

%

%% file: Content/SamplingAngles.tex
\section{Models and Preliminaries for Array and Angular Domains}\label{sec:ArrayAngularDomains}
\subsection{Antenna array model}
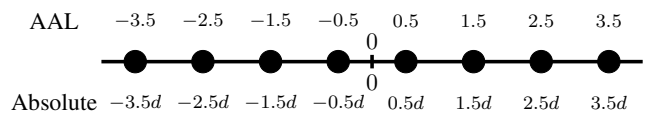
\begin{figure}[h!]
  \centering
  \resizebox{\linewidth}{!}{
  \if\single1
\input{Figures/ArrayExample_SingleColumn.tikz}
\else
\input{Figures/ArrayExample.tikz}
\fi
  
  }
  \caption{\gls{ULA} example with $N = 8$ elements, with their absolute location $x$ and \gls{AAL} $x' = x/d$. } 
  \label{fig:ArrayExample}
\end{figure}

We assume to sense the environment over a single angular dimension - the azimuth - with a \gls{ULA}  
by transmitting or receiving focused beams of narrowband signals with wavelength $\lambda$.
Let $N$ be the number of array elements that are equally spaced $d$ apart, whose absolute positions are defined by the set 
\if\single1
\begin{align}
\mathcal{X} = \{x_0, ..., x_{N-1}\}  \; \text{, with }
\label{eq:ArrayAntennasLocations}
x_n = \left(-\frac{N-1}{2} + n \right) d \;,\;\; n = 0, ..., N-1 \;.
\end{align}
\else
\begin{align}
\mathcal{X} &= \{x_0, ..., x_{N-1}\}  \; \text{, with }
\nonumber \\
\label{eq:ArrayAntennasLocations}
x_n &= \left(-\frac{N-1}{2} + n \right) d \;,\;\; n = 0, ..., N-1 \;.
\end{align}
\fi
We then define the \gls{LAD} as $\ell = \frac{d\sin \theta}{\lambda}$, and the \gls{AAL} $x'$ as a virtual position axis with unitary antenna spacing, i.e., $x' = x/d$ and $x'_n = x_n/d$. See Fig.~\ref{fig:ArrayExample} for an illustrative example with $N = 8$. 

The signal at the $n$-th element of such an array with a unitary amplitude planar wave at incident  angle $\theta$ can be seen as the sampled response in the \gls{AAL} domain 
\begin{equation}
{a}_n(\ell) 
= e^{j{2\pi}x'_n \ell } \; .
\label{eq:ArrayLADResponse}
\end{equation}
\begin{remark}\label{remark:UnitaryLAD}
The \gls{LAD} axis has a physical representation in the interval $[-{d}/{\lambda}, {d}/{\lambda}]$ of extension $2d/\lambda$, corresponding to incident angles in $[-\pi/2, \pi/2]$. However, the \gls{LAD} can assume any real value in $(-\infty, +\infty)$ due to the periodic array response (see Sec.~\ref{subsec:ArrayLimits}).
\end{remark}
From Eq.~\eqref{eq:ArrayLADResponse}, given an arbitrary incident integral of planar waves, describing the sparse angular response (or image) of the scenario, one can write the signal at the $n$-th antenna as
\begin{equation}
{a}_n 
= \int_{-\frac{d}{\lambda}}^{\frac{d}{\lambda}} L(\ell) e^{j{2\pi}x'_n \ell } d\ell \; ,
\label{eq:FourierDuality}
\end{equation}
where $L(\ell)$ is the coefficient of the incident planar wave with \gls{LAD} $\eta$.
Eq.~\eqref{eq:FourierDuality} establishes a Fourier duality between the \gls{LAD} $\ell$ and the \gls{AAL} $x'$. This means that one domain can be obtained as the (inverse) Fourier transform of the other. 
%
Increasing element spacing $d$ has benefits of antenna diversity and angular resolution, due to larger array aperture~\cite{steinberg1976principles}, but at the cost of increased possibility of aliases in the angular domain.
This can be explained by the fact that antenna elements in a \gls{ULA} uniformly sample the signal in the \gls{AAL} domain at the antenna locations, creating replica every unitary shifts in its Fourier transform~\cite{pinsky2008introduction}, i.e., the \gls{LAD} domain.

Leveraging the sampling theorem~\cite{shannon1949communication}, if one wants to avoid aliases in the \gls{LAD} over the interval $[-{d}/{\lambda}, {d}/{\lambda}]$,
one has to sample the \gls{AAL} axes by keeping a gap between the \gls{ULA}'s antennas, satisfying the following (well-known) condition
\begin{align}
{\underbrace{1}_\text{\gls{AAL} Sampling frequency}} \geq \underbrace{\frac{2d}{\lambda}}_{\text{\gls{LAD} extension}} \;  \Leftrightarrow
\; \; d \leq \frac{\lambda}{2} \;.
\label{eq:SamplingTheoremAAL}
\end{align}
Note that if the \gls{LAD} aperture of interest is reduced to $2dp/\lambda$, with $p<1$, the \gls{AAL} domain could be sampled more coarsely, i.e. $d \leq \lambda/(2p) \geq \lambda/2$. In the special case of $d = \lambda/2$, \gls{LAD} replica are generated at unitary period. Therefore, one could only consider $[-\pi/2, \pi/2)$ or $(-\pi/2,\pi/2]$ as angular intervals without aliasing.

\subsection{Impact of array aperture}\label{subsec:ArrayLimits}
For a \gls{ULA} with $N$ antenna elements, the array total aperture in the \gls{AAL} domain induces a low pass frequency effect in the \gls{LAD} domain, due to the Fourier duality discussed earlier.  
To maximize the \gls{ULA}'s response on a given \gls{LAD} $\ell$, the array matched beamformer can be applied to get the array's \textit{\gls{LAD} response}
\begin{equation}
L_N(\ell) =
\frac{1}{N} \sum_{n=0}^{N-1} a_n e^{-j 2 \pi x'_n \ell} \;,
\label{eq:RxBeamformer}
\end{equation}
where $a_n$ is again the signal at the $n$-th antenna element. 
For analog beamforming, the array focuses on a different \gls{LAD} $\ell$ (thus sampling along a corresponding direction in the angular domain) by applying weights $w_n(\ell) = e^{-j 2 \pi x'_n \ell}$ as in \eqref{eq:RxBeamformer}.
From Remark \ref{remark:UnitaryLAD}, one can notice that $L'_N(\ell), \ell \notin [-{d}/{\lambda}, {d}/{\lambda}]$ focuses on a \gls{LAD} that does not represent a true physical angle.

\begin{lemma}\label{lemma:PlanarWaveResponse}
The \gls{ULA} \gls{LAD} response \eqref{eq:RxBeamformer} of a single incident planar wave at \gls{LAD} $\eta$ is
\begin{equation}
L_N(\ell, \eta) = \frac{\sin \left( \pi N (\ell - \eta)  \right)}{N \sin \left( \pi  (\ell - \eta) \right)}
= D_N \left( \ell - \eta \right)
 \; .
\label{eq:PlanarWaveResponse}
\end{equation}
\end{lemma}
\begin{proof}
\if\onappendix1
See Appendix \ref{app:PlanarWaveResponse}. $D_N \left( \ell \right) {\equiv} \sin(\pi N \ell)/(N \sin (\pi \ell))$ is the Dirichlet kernel of order $N$.
\else
$D_N \left( \ell \right) {\equiv} \sin(\pi N \ell)/(N \sin (\pi \ell))$ is the Dirichlet kernel of order $N$. The complete proof can be found in Appendix A of the supplementary material.
\fi

\end{proof}
The main lobe half-width and period of the Dirichelt kernel in~\eqref{eq:PlanarWaveResponse} correspond to the multi-target resolution of $1/N$ and unitary aliasing, respectively, of \gls{LAD} estimation algorithms. 

\begin{remark}
Amplitude scaling coefficients could be applied at each antenna element to suppress side-lobes at the price of beam gain. 
In particular, \eqref{eq:RxBeamformer} can work with any arbitrary, hence optionally scaled, complex signal at each antenna element. 
\end{remark}

We define now an auxiliary \gls{LAD} response
\begin{equation}
 L'_N(\ell) = L_N(\ell)e^{-j2\pi\frac{N-1}{2}\ell} \; ,
\label{eq:RxBeamformerShifted}
\end{equation}
where we leveraged the Fourier duality to translate the \gls{AAL} axes such that the first antenna element is at \gls{AAL} equal to zero.  This allows to write the following Lemma.
\begin{lemma}\label{lemma:NAFPeriodicity}
The auxiliary \gls{LAD} response of a \gls{ULA} defined in \eqref{eq:RxBeamformer} is a periodic function given by
\begin{equation}
L'_N\left(\ell - k  \right) = L'_n(\ell) \;,
\text{ with } k \in \mathbb{Z} \;.
\label{eq:LADPeriodicity}
\end{equation}

\end{lemma}
\begin{proof}
\if\onappendix1
The complete proof is in Appendix \ref{app:NAFPeriodicity}.
\else
The proof is based on expanding $L'_N\left(\ell - k \right)$, using its definition \eqref{eq:RxBeamformer}. The complete proof can be found in Appendix B of the supplementary material.
\fi
\end{proof}
\begin{remark}
The array extension in \gls{AAL} is in $\left[-\frac{N-1}{2}, \frac{N-1}{2}\right]$. After the auxiliary translation in \gls{AAL}, the extension is shifted to $\left[0, N-1\right]$.
\label{remark:AalExtension}
\end{remark}

%% file: Figures/ArrayExample_SingleColumn.tikz.tex
\begin{tikzpicture}[antennastyle/.style={circle,draw,fill=black,minimum size=0.6}]

\def\lw{1.5}
\def\height{0.6}

\draw[-, line width=\lw]  (-0.5, 0) -- (7.5,0); 
\node at (-1.5,-\height) {\footnotesize Absolute $x$};
\node at (-1.5,\height)  {\footnotesize \gls{AAL} $y$};
\draw[-, line width=\lw]  (3.5, -0.1) -- (3.5,0.1);
\node at (3.5, -0.3)  {\footnotesize $0$};
\node at (3.5, +0.3)  {\footnotesize $0$};

\foreach \x in {0,...,7}
	{\node[antennastyle] at (\x,0) {};
	\pgfmathsetmacro\position{(-3.5 + \x)}
	\node at (\x,-\height) {\footnotesize $\position d$ };
	\node at (\x,\height) {\footnotesize $\position $ };}

\end{tikzpicture}

%% file: Figures/ArrayExample.tikz.tex
\begin{tikzpicture}[antennastyle/.style={circle,draw,fill=black,minimum size=0.6}]

\def\lw{1.5}
\def\height{0.6}

\draw[-, line width=\lw]  (-0.5, 0) -- (7.5,0); 
\node at (-1.2,-\height) {Absolute};
\node at (-1.2,\height)  {\gls{AAL}};
\draw[-, line width=\lw]  (3.5, -0.1) -- (3.5,0.1);
\node at (3.5, -0.3)  {$0$};
\node at (3.5, +0.3)  {$0$};

\foreach \x in {0,...,7}
	{\node[antennastyle] at (\x,0) {};
	\pgfmathsetmacro\position{(-3.5 + \x)}
	\node at (\x,-\height) {\footnotesize $\position d$ };
	\node at (\x,\height) {\footnotesize $\position$ };}

\end{tikzpicture}

%% file: Content/AngularReconstruction.tex
\section{Angular Sampling and Reconstruction}\label{sec:AngularReconstruction}
In this section, we propose a method to sample and reconstruct the angular domain (SARA) using analog \gls{ULA} systems.
With \gls{SARA}, the reconstruction is ``loss-less'' in the sense that it allows the perfect angular response reconstruction as if noise-less infinite scans at every angle were performed using the considered \gls{ULA}. 
The Nyquist-Shannon sampling theorem~\cite{shannon1949communication} states that a sufficient condition to completely determine and reconstruct a generic time signal with (positive and negative) bandwidth $B$ is to sample it uniformly at points $1/B$ apart.
Then, the original signal can be perfectly reconstructed into $x_r(t)$ from its infinite samples $x(nB^{-1})$ by low pass filtering them, getting the well known Whittaker-Shannon interpolation formula
\begin{equation}
x_r(t) = \sum_{n = -\infty}^{+\infty} x(nB^{-1}) \; \text{sinc}\left( B \left( t - \frac{n}{B} \right) \right) \;,
\label{eq:Whittaker-Shannon-Interpolation}
\end{equation}
where $\text{sinc}(x) = \sin(\pi x)/(\pi x)$.
We frame the problem of sampling and reconstructing a \gls{ULA} angular response by working in the \gls{AAL} and \gls{LAD} domains, leveraging their Fourier duality.
One can notice from \eqref{eq:ArrayAntennasLocations} that the \gls{AAL} domain can avoid aliasing and have equally spaced samples if the replica generated from sampling are $N$ apart from each other.
Therefore, we can sample the \gls{ULA}'s \gls{LAD} response with frequency $N$, obtaining   $L(nN^{-1})$. 
Applying \eqref{eq:Whittaker-Shannon-Interpolation}, the reconstructed \gls{LAD} response ($B = N^{-1}$) is obtained as
\begin{equation}
R_N(\ell) =  
\sum_{n = - \infty}^{+\infty}
L_N \left( \frac{n}{N} \right)
\text{sinc} \left( N \left( \ell - \frac{n}{N} \right) \right)\;.
\label{eq:InfiniteReconstructionSummation}
\end{equation}
Here, we make the auxiliary \gls{LAD} response derivation explicit, since the periodic behavior of the \gls{LAD} response $L_N(\ell)$ is not trivial as seen in Lemma~\ref{lemma:NAFPeriodicity}.
\if\single1
\begin{align}
R'_N(\ell) =  
\sum_{n = - \infty}^{+\infty}& \left( L'_N \left( \frac{n}{N} \right)
\text{sinc} \left( N \left( \ell - \frac{n}{N} \right) \right) e^{-j2\pi \frac{N-1}{2} \left( \ell - \frac{n}{N}  \right)} \right)  \;,
\end{align}
\else
\begin{align}
R'_N(\ell) =  
\sum_{n = - \infty}^{+\infty}& \left( L'_N \left( \frac{n}{N} \right)
\text{sinc} \left( N \left( \ell - \frac{n}{N} \right) \right) \cdot \right.
\nonumber \\
 &\left.\cdot\,e^{-j2\pi \frac{N-1}{2} \left( \ell - \frac{n}{N}  \right)} \right)  \;,
\end{align}
\fi
where the exponential term inside the summation is making sure that the reconstruction is centered at \gls{AAL} equal to $\frac{N-1}{2}$, as seen in Remark \ref{remark:AalExtension}. One can revert the transformation from the auxiliary \gls{LAD} response to the \gls{LAD} response, having $R_N(\ell) = R'_N(\ell) e^{j2\pi \frac{N-1}{2}\ell}$. 
Given the replica generated according to Lemma \ref{lemma:NAFPeriodicity}, one could define the minimum set of indexes without replica as $\mathcal{N}_N = \{n \in \mathbb{Z} : -\frac{N}{2} \leq n < \frac{N}{2} \}$, and just sample them as follows 
\if\single1
\begin{align}
R'_N(\ell) &=  
\sum_{n \in \mathcal{N}_N} \sum_{k = -\infty}^{+\infty}
\text{sinc} \left( N \left( \ell - \frac{n + kN}{N} \right) \right) 
L'_N \left( \frac{n + kN}{N} \right) e^{-j2\pi \frac{N-1}{2} \left( \ell - \frac{n}{N}  \right)}=
\nonumber \\
&= \sum_{n \in \mathcal{N}_N} 
L'_N \left( \frac{n}{N} \right) \cdot \sum_{k = -\infty}^{+\infty} e^{-j2\pi \frac{N-1}{2} \left( \ell - \frac{n+kN}{N} \right)}  \text{sinc} \left( N\left( \ell - \frac{n+kN}{N} \right) \right) \;.
\label{eq:InterpolationSincStep1}
\end{align}
\else
\begin{align}
R'_N(\ell) &=  
\sum_{n \in \mathcal{N}_N} \sum_{k = -\infty}^{+\infty}
\text{sinc} \left( N \left( \ell - \frac{n + kN}{N} \right) \right) \cdot \nonumber \\
&\cdot
L'_N \left( \frac{n + kN}{N} \right) e^{-j2\pi \frac{N-1}{2} \left( \ell - \frac{n}{N}  \right)}=
\nonumber \\
&= \sum_{n \in \mathcal{N}_N} 
L'_N \left( \frac{n}{N} \right) \cdot \sum_{k = -\infty}^{+\infty} e^{-j2\pi \frac{N-1}{2} \left( \ell - \frac{n+kN}{N} \right)} \cdot
\nonumber \\
&\cdot \text{sinc} \left( N\left( \ell - \frac{n+kN}{N} \right) \right) \;.
\label{eq:InterpolationSincStep1}
\end{align}
\fi
\if\onappendix1
We show in Appendix \ref{app:InfiniteSinc}
\else
We show in Appendix D of the supplementary material
\fi
how one can simplify the infinite summation part of \eqref{eq:InterpolationSincStep1}, obtaining
\if\single1
\begin{align}
R'_N(\ell) &= 
\sum_{n \in \mathcal{N}_N}
L'_N \left( \frac{n}{N} \right) D_N\left( \ell - \frac{n}{N} \right)
e^{-j2\pi \frac{N-1}{2} \left( \ell - n\frac{\lambda}{N} \right)}
=
\sum_{n \in \mathcal{N}_N}
L'_N \left( \frac{n}{N} \right) D'_N\left(  \ell - \frac{n}{2dN}  \right)
\; ,
\label{eq:InterpolationGenericFinalAuxiliary}
\end{align}
\else
\begin{align}
R'_N(\ell) &= 
\sum_{n \in \mathcal{N}_N}
L'_N \left( \frac{n}{N} \right) D_N\left( \ell - \frac{n}{N} \right)\cdot
\nonumber \\
&\cdot 
e^{-j2\pi \frac{N-1}{2} \left( \ell - n\frac{\lambda}{N} \right)}
= \nonumber \\
&=
\sum_{n \in \mathcal{N}_N}
L'_N \left( \frac{n}{N} \right) D'_N\left(  \ell - \frac{n}{2dN}  \right)
\; ,
\label{eq:InterpolationGenericFinalAuxiliary}
\end{align}
\fi
which is a convolution of the sampled sequence with Dirichlet kernels with a linear phase component, defined as $D_N'(\ell) = D_N(\ell)e^{-j \pi (N-1)\ell}$. The derivation of the reconstructed \gls{LAD} response can be  obtained similarly, with the absence of any linear phase shift in the Dirichlet kernel, making it equivalent to a trigonometric interpolation~\cite{jackson1913accuracy} of the samples taken uniformly in the \gls{LAD} domain 
\begin{align}
R_N(\ell) 
&=
\sum_{n \in \mathcal{N}_N}
L_N \left( \frac{n}{N} \right) D_N\left(  \ell - \frac{n}{N}  \right) \; .
\label{eq:InterpolationGenericFinal}
\end{align}


We define the following continuous functions corresponding to the angular scans and limited Dirichlet kernel with shifted phase, respectively
\begin{align}
\overline{L}'_N(\ell) &= \sum_{n \in \mathcal{N}_N} L'_N(\ell) \; \delta \left( \ell  - \frac{n}{N} \right) \;,
\label{eq:LADPeriodical}
\\
\overline{D}'_N \left( \ell \right) &=
\begin{cases}
D'_N \left( \ell \right) & \text{ if } |\ell| \leq 0.5 \\
0         & \text{ otherwise}
\end{cases} \;.
\label{eq:LimitedDirichlet}
\end{align}
\begin{theorem}
\label{theo:circular}
The reconstructed auxiliary \gls{LAD} response in \eqref{eq:InterpolationGenericFinalAuxiliary} is equivalent to a circular convolution of unitary period of the two functions $\overline{L}'_N(\ell)$ and $\overline{D}'_N(\ell)$.
\end{theorem}
\begin{proof}
Using the definition in \eqref{eq:LADPeriodical}, one can reshape \eqref{eq:InterpolationGenericFinal} into
\if\single1
\begin{align}
R'_N(\ell) &= \int_{-\infty}^{+\infty} \overline{L}'(\eta) \overline{D}'_N (\ell - \eta)  d \eta =
\overline{L}'(\ell) * \overline{D}'_N(\ell) \;.
\label{eq:LADLinearConvolution}
\end{align}
\else
\begin{align}
R'_N(\ell) &= \int_{-\infty}^{+\infty} \overline{L}'(\eta) \overline{D}'_N (\ell - \eta)  d \eta =
\nonumber \\
&= 
\overline{L}'(\ell) * \overline{D}'_N(\ell) \;.
\label{eq:LADLinearConvolution}
\end{align}
\fi
Then, given the auxiliary \gls{LAD} response periodicity, see Lemma~\ref{lemma:NAFPeriodicity}, one can write the linear convolution~\eqref{eq:LADLinearConvolution} as a periodical convolution~\cite{priemer1991introductory} with unitary period as
\begin{align}
R'_N(\ell) &= \int_{-0.5}^{0.5} \overline{L}'(\eta) \sum_{k = -\infty} ^ {+ \infty} \overline{D}'_N  (\ell - \eta - k) d \eta \;.
\end{align}
The previous equation holds since all non-zero elements of $L'(\ell)$ and $\overline{D}'_N(\ell)$ lie in $[-0.5, 0.5)$, given their definitions in~\eqref{eq:LADPeriodical}-\eqref{eq:LimitedDirichlet}. Therefore, as shown in~\cite{priemer1991introductory}, the periodical convolution is also a circular convolution with unitary period of the two aperiodic functions $\overline{L}'_N(\ell)$ and $\overline{D}'_N(\ell )$.
\end{proof}
In practical applications, one is not interested in all infinite \gls{LAD} values but  a fine grid in the \gls{LAD} interval of interest. Therefore, the available samples $L'_N(nN^{-1}), n \in \mathcal{N}_N$, must be up-sampled by a factor $U$, getting all $R'_N(u(NU)^{-1}), u \in \mathcal{N}_{NU}$.
We define the vector $\mathbf{\overline{l}}'$, whose elements are $\mathbf{\overline{l}}'_{n-\underline{N}} = L'_N(nN^{-1})$, with $n \in \mathcal{N}_N$ and $\underline{N}$ being equal to the infimum of $\mathcal{N}_N$.
One can create the following vectors of $NU$ elements, spanning the period of the continuous functions defined in \eqref{eq:LADPeriodical} and \eqref{eq:LimitedDirichlet}, sampled at $\ell = u/(NU)$ 
\begin{align}
\mathbf{l}'_{u} &= \begin{cases}
\mathbf{\overline{l}}'_n & \text{ if } u = Un \text{ , } n = 0, ..., N-1 \\
0                 & \text{ otherwise} 
\end{cases} 
\label{eq:UpsampledResponse} \; ,\\ 
\mathbf{d}'_{u} &= \overline{D}'_N \left( \frac{u}{NU} \right) = D'_N \left( \frac{u}{NU} \right) \;,
\label{eq:UpsampledDirichlet}
\end{align}
where $u \in \{u \in \mathbb{N}_0 : u < NU\} = \mathcal{U}_{NU}$.
\begin{theorem}\label{theo:LR}
The up-sampled auxiliary \gls{LAD} response can be obtained with a circularly sampled convolution in the \gls{LAD} domain of the vectors $\mathbf{l}'$ and $\mathbf{d}'$ of Equations \eqref{eq:UpsampledResponse}-\eqref{eq:UpsampledDirichlet}.
\end{theorem}
\begin{proof}
Theorem \ref{theo:circular} states that the desired reconstructed \gls{LAD} response can be obtained as a circular convolution between $\overline{L}'_N(\ell)$ and $\overline{D}'_N(\ell)$. If the sampling theorem is satisfied, one could perfectly reconstruct the continuous \gls{LAD} signal, or (up-)sample it.
Therefore, the reconstructed up-sampled auxiliary \gls{LAD} response can be written as the following circular convolution of $\mathbf{l}'$ with $\mathbf{d}'$
\begin{equation}
\mathbf{r}'_u = \sum_{v \in \mathcal{U}} \mathbf{l}'_v \mathbf{d}'_{ \text{mod}_{NU} (u-v) }
\text{ with } u \in \mathcal{U}_{NU} \;.
\label{eq:LR}
\end{equation}
\end{proof}
Given the convolution of $N$ non-zero samples of $\mathbf{l}'$ with $NU$ samples of $\mathbf{d}'$, the \gls{LR} technique has complexity $\mathcal{O}(UN^2)$. 
\begin{lemma}\label{lemma:FftDirichlet}
The inverse \gls{DFT} transforms with $NU$ elements ($\text{IDFT}_{NU}$) of $\mathbf{l}'$ and $\mathbf{d}'$ can be written, respectively, as follows
\begin{align}
\mathbf{l}'_A &= \text{IDFT}_{NU} \left( \mathbf{l}' \right) =  \left[ \text{IDFT}_N\left(\mathbf{\overline{l}'} \right), ..., \text{IDFT}_N\left(\mathbf{\overline{l}'} \right) \right]
\label{eq:ResponseDFT} \; ,
\\
\mathbf{d}'_A &= \text{IDFT}_{NU} \left( \mathbf{d}' \right) = \left[ \mathbf{1}_N, \mathbf{0}_{NU-N} \right] \;,
\label{eq:DirichletDFT}
\end{align}
where $\text{IDFT}_N\left(\mathbf{\overline{l}'} \right)$ in~\eqref{eq:ResponseDFT} is with $N$ elements and it is repeated $U$ times sequentially.
\end{lemma}
\begin{proof}
The complete proof can be found in Appendix~\ref{app:IdftDerivation}.
\end{proof}

\begin{theorem}\label{theo:AR}
The up-sampled reconstructed auxiliary \gls{LAD} response can be obtained from $\mathbf{\overline{l}'}$ 
as  
\begin{equation}
\mathbf{r'} =  \text{DFT}_{NU} \left( \left[
\text{IDFT}_{N} \left( \mathbf{\overline{l}'} \right), 
\mathbf{0}_{NU-N}
\right]
\right) \;,
\label{eq:AR}
\end{equation}
corresponding to a \gls{DFT} interpolation of the auxiliary sampled response.
\end{theorem}
\begin{proof}
Since circular convolution in one domain is equivalent to an element-wise multiplication of the IDFT/\gls{DFT} transforms~\cite{priemer1991introductory}, the \gls{AR} of \eqref{eq:AR} can be obtained from \eqref{eq:LR} by applying IDFT to both $\mathbf{l}'$ and $\mathbf{d}'$, applying element-wise multiplication (Hadamard product), and finally obtaining the result in the \gls{LAD} domain with the \gls{DFT} operation
\if\single1
\begin{align}
\mathbf{r'} &= \text{DFT}_{NU} \left( 
\text{IDFT}_{NU} \left( \mathbf{{l}'} \right) \odot 
\text{IDFT}_{NU} \left( \mathbf{{l}'} \right)
\right) =
\nonumber \\
& \overset{\eqref{eq:ResponseDFT},\eqref{eq:DirichletDFT}}{=}\text{DFT}_{NU} \left( 
\left[ \text{IDFT}_N\left(\mathbf{\overline{l}'} \right), ... \right] \odot 
\left[ \mathbf{1}_N, \mathbf{0}_{NU-N} \right]
\right) = 
\text{DFT}_{NU} \left( \left[
\text{IDFT}_{N} \left( \mathbf{\overline{l}'} \right), 
\mathbf{0}_{NU-N}
\right]
\right) \;. \nonumber
\end{align}
\else
\begin{align}
\mathbf{r'} &= \text{DFT}_{NU} \left( 
\text{IDFT}_{NU} \left( \mathbf{{l}'} \right) \odot 
\text{IDFT}_{NU} \left( \mathbf{{l}'} \right)
\right) =
\nonumber \\
& \overset{\eqref{eq:ResponseDFT},\eqref{eq:DirichletDFT}}{=}\text{DFT}_{NU} \left( 
\left[ \text{IDFT}_N\left(\mathbf{\overline{l}'} \right), ... \right] \odot 
\left[ \mathbf{1}_N, \mathbf{0}_{NU-N} \right]
\right) = 
\nonumber \\
&=
\text{DFT}_{NU} \left( \left[
\text{IDFT}_{N} \left( \mathbf{\overline{l}'} \right), 
\mathbf{0}_{NU-N}
\right]
\right) \;. \nonumber
\end{align}
\fi
\end{proof}

Accordingly, the resulting complexity is $\mathcal{O}(NU \log (NU))$ due to the final \gls{DFT} operation, that however is applied to a vector of only $N$ non-zero elements.

%% file: Content/ULAReconstruction.tex
\section{Practical Applications of \gls{SARA} }\label{sec:ULAReconstruction}
\subsection{Omni-directional transmitter and directional receiver}\label{subsec:DirectionalRx}
\begin{algorithm}[ht]
\begin{algorithmic}[1]
\Procedure{Scan and Interpolate}{}
\State{$\mathbf{\overline{l}} \leftarrow$ scan $L_N(\eta)$~\eqref{eq:RxBeamformer}, with $\eta \in \mathcal{L}_N^{(\text{Rx})}$ from~\eqref{eq:AnglesToScanOnlyReceiver}}
\State{$\mathbf{\overline{l}'} \leftarrow$ sampled auxiliary response~\eqref{eq:RxBeamformerShifted}}
\State{$\mathbf{l}' \leftarrow$ up-sample $\mathbf{\overline{l}'}$~\eqref{eq:UpsampledResponse} only with \gls{LR}}
\State{$\mathbf{r}' \leftarrow $ Up-sampled \gls{LAD} response with \gls{LR} or \gls{AR}, using Theorem \ref{theo:LR} or \ref{theo:AR}, respectively}
\State{$\mathbf{r} \leftarrow $ Non-auxiliary response, inverting~\eqref{eq:RxBeamformerShifted}}
\State \Return{$\mathbf{r}$}
\EndProcedure
\end{algorithmic}
\caption{\gls{LAD} reconstruction with $N$ elements \gls{ULA}. \label{alg:OnlyReceiverReconstruction}}
\end{algorithm}
Assuming that the scenario is already illuminated by another device, 
we want to reconstruct the \gls{LAD} response by up-sampling a finite angular scan by a factor $U$.
The reconstruction procedure to get the reconstructed (non-auxiliary) \gls{LAD} response $\mathbf{r}$ is sketched in Algorithm \ref{alg:OnlyReceiverReconstruction}. 
In particular, to obtain the \gls{LAD} response without the auxiliary \gls{AAL} translation, one can apply~\eqref{eq:RxBeamformerShifted} to its available samples $\mathbf{\overline{l}}$, obtaining the sampled auxiliary response. Then, the auxiliary response can be reconstructed with Theorems~\ref{theo:circular}, \ref{theo:LR}, or \ref{theo:AR}, and then~\eqref{eq:RxBeamformerShifted} can be inverted to get the desired reconstructed \gls{LAD} response. 
The \gls{LAD} to be scanned are 
\begin{equation}
\mathcal{L}_{N}^{(\text{Rx})} = \{ n N^{-1} : n \in \mathcal{N}_N \} \; .
\label{eq:AnglesToScanOnlyReceiver}
\end{equation}
%
%
%
%
\subsection{Different number of \gls{LAD} samples}\label{subsec:DifferentNumberScans}
So far, this work determined that $N$ angles must be scanned to fully reconstruct the angular response of a \gls{ULA} with $N$ elements. In case a number of angular scans $\overline{N} \neq N$ is available, previous considerations can be modified to make the most out of the available scans.
In case more scans are available, i.e. $\overline{N} > N$, one could have a finer sampling of the \gls{LAD} domain, thus sample angles in $\mathcal{L}_{\overline{N}}^{\text{Rx}}$. When reconstructing the signal, one should note from \eqref{eq:LADPeriodicity} that the replica are still generated by the $N$ elements of the \gls{ULA}. Therefore, steps similar to \eqref{eq:InterpolationGenericFinal} and \if\onappendix1Appendix \ref{app:InfiniteSinc} 
\else
Appendix C of the supplementary material
\fi
could be applied with an $N$ elements \gls{ULA}, but sampling the angular domain every $\overline{N}$, getting the reconstruction formula for $\overline{N} \neq N$
\begin{equation}
L'(\ell) = \frac{N}{\overline{N}}
\sum_{l \in \mathcal{L}_{\overline{N}}}
L_N \left( l \right) D_N\left( \left( \ell - l \right)  \right) \; .
\label{eq:InterpolationUpsamples}
\end{equation}
The factor $N/\overline{N}$ comes from the different normalization factors in \if\onappendix1\eqref{app_eq:SincTransform1} of the 
\else
Equation (45) in the supplementary material, corresponding to the
\fi
rectangle with reduced aperture $\overline{N}$, whereas the train of impulses retains the same unitary periodicity.


On the other hand, a loss of information is experienced in the reconstructed \gls{LAD} response if less angles than the necessary amount can be scanned. This can happen due to the resource shortage in the wireless system, or due to the analog beam codebook having a reduced number of total beams that can be stored. However, if the full non-aliased angular unitary aperture is not of interest, one could focus the available $\overline{N}$ scans in the directions of interest $\mathcal{D} \subset \mathcal{L}_N^{\text{(Rx)}}$, still sampling the \gls{LAD} domain with period $1/N$ and then interpolating, only in the interval spanned by $\mathcal{D}$. The missing scans needed for the \gls{LAD} reconstruction, see \eqref{eq:LADPeriodical}, can be set to zero, i.e. $L'_N(\ell) = 0, \forall n \in \mathcal{L}_N^{\text{(Rx)}} \setminus \mathcal{D}$. 
Minor losses in terms of the reconstructed \gls{LAD} response's distortion come from not probing the full \gls{LAD} domain. Note that the response of an angular impulsive scatterer is spread over the full \gls{LAD} domain aliasing period due to the finiteness of the array aperture. However, this approach allows to preserve the resolution in the directions of interest.
Alternatively, if the full \gls{LAD} domain aliasing period is of interest, one could reduce the number of operating antennas to $\overline{N}$ contiguous elements and perform the reduced scan accordingly. This results in a reconstructed \gls{LAD} response with lower resolution, due to the reduced aperture of the operating array.
The two approaches above can be trivially combined in case the number of necessary scans is not sufficient to scan a reduced angular range $\mathcal{D}$.


\subsection{Directional transmitter and receiver}\label{subsec:DirectionalTxRx}
For sensing applications, we may have a transmit array illuminating the environment where beamforming can be applied both at the transmitter and at the receiver side. Their combined effect can be seen as a multiplication of their responses in the \gls{LAD} domain, thus a convolution in the \gls{AAL} domain. 
This allows us to unify our dissertation with the literature on sum co-array~\cite{hoctor1990unifying}, where the achievable Point Spread Function of a transmit and a receive array is defined by their sum co-array. 
The sum co-array is a virtual array structure defined as
the set of pairwise sums of the transmit and receive elements'
locations. 
Note that the convolution in the \gls{AAL} domain of discrete sequences, whose non-zero elements are at the transmitter's and receiver's locations $\mathcal{X}$, is located at the sum co-array's virtual locations.
In the particular case of a full duplex \gls{ULA} with $N$ elements, it is straightforward to show that the sum co-array is a \gls{ULA} with $2N - 1$ elements. Therefore, one should operate as if a $2N-1$ \gls{ULA} were available to  perform \gls{SARA}. Note the almost doubled resolution, which is due to the joint efforts of transmitter and receiver.

\subsection{Extension to \gls{URA}}\label{subsec:URAExtension}
Similar techniques can be extended to \gls{2D} arrays to obtain angular estimates in both azimuth and elevation. In particular, one could generalize the one-dimensional \gls{SARA} considerations tailored for \gls{ULA} to \gls{URA}.
We assume an $N\text{x}M$ \gls{URA} structure with vertical (z axis) and horizontal (x axis) antenna spacing corresponding to $d_1$ and $d_2$, respectively. Its angular response at elevation $\phi$ and azimuth $\theta$ can be written as~\cite{van2004optimum}
\begin{align}
L_{N,M}(\phi, \theta) &= \sum_{n=0}^{N-1}\sum_{m=0}^{M-1} a_{n,m} 
e^{-j 2 \pi \left( z'_n \frac{d_1}{\lambda} \sin\left( \phi \right) +
x'_m \frac{d_2}{\lambda} \frac{\sin\left( \theta \right)}{ \cos(\phi)} \right)},
\label{eq:URARxBeamformer}
\end{align}
where $z'_n$ is the $n$-th antenna location in the vertical \gls{AAL} axis. We recall the constraint of $|\phi|+|\theta| \leq \pi/2$ in spherical coordinates.
From~\eqref{eq:URARxBeamformer} we notice that the vertical phase shift depends on a vertical \gls{LAD} $\eta = \frac{d_1}{\lambda}\sin(\phi)$, whereas the horizontal phase shift on a horizontal \gls{LAD} of $\ell = \frac{d_2}{\lambda}\sin(\theta) / \cos(\phi)$.
Since the horizontal \gls{LAD} depends on the elevation, the physical azimuth and elevation cannot be decoupled and are obtained from the \gls{LAD} dimensions as 
\begin{align}
\phi(\eta) &= \sin^{-1}\left(\frac{\lambda}{d_1} \eta \right) \; ,
\label{eq:URAElevation}
\\
\theta(\ell, \eta) &= \sin^{-1}\left(   \frac{\lambda}{d_2} \cos(\phi(\eta))\ell\right) \; .
\label{eq:URAAzimuth}
\end{align}
With procedures similar to the ones shown in Section \ref{sec:AngularReconstruction}, one can demonstrate that these two \gls{LAD} dimensions are the (orthogonal) Fourier dual of the vertical and horizontal \gls{AAL} axes. Accordingly, all theorems in Section~\ref{sec:AngularReconstruction} can be extended for the \gls{2D} case. Therefore, the full \gls{2D} \gls{LAD} response can be obtained by performing $NM$ scans along directions given by the vertical and horizontal \gls{LAD} sets given by
\begin{equation}
\mathcal{L}_{N,M}^{(Rx)} = \{ \eta=nN^{-1}, \ell = mM^{-1} : n \in \mathcal{N}_N, m \in \mathcal{N}_M \}\;,    
\label{eq:URA_Angles}
\end{equation}
where~\eqref{eq:URAElevation}-\eqref{eq:URAAzimuth} can be used to determine the corresponding elevation and azimuth angles. 
The \gls{2D} angular reconstruction can be obtained by convolving the scan with the following \gls{2D} Dirichlet kernel
\begin{equation}
D_{N,M}(\eta, \ell) =
D_N\left( \eta  \right) \cdot
D_M\left( \ell  \right) \; ,
\end{equation}
whose inputs are the vertical and horizontal \glspl{LAD}, respectively.
Accordingly, Theorems \ref{theo:circular}, \ref{theo:LR} and \ref{theo:AR} can be extended to cover the \gls{2D} angular space. 
Regarding the sensing applications mentioned in Subsection \ref{subsec:DirectionalTxRx}, the role of the sum co-array structure still holds when a transmitting and a receiving \gls{URA} operate together.
In case of a co-located \gls{URA} with $N\text{x}M$ elements, one still has a $(2N-1)\text{x}(2M-1)$ \gls{URA} structure determining the performance and features of \gls{SARA}.

%% file: Content/Simulation.tex
\section{Numerical Experiments}\label{sec:Simulation}
\subsection{Simulation Setup}\label{subsec:simScenario}
In our simulation experiments, a single terminal sensing scenario with full duplex capabilities is considered, equivalent to mono-static radar. The array consists of a horizontal $N$-element analog \gls{ULA} with a corresponding sum co-array of $2N{-}1$ elements. 
The antennas experience the same \gls{AWGN} power $\sigma_n^2$ and are separated by $d = \lambda/2$, making the non-aliased unitary period of the \gls{LAD} axes correspond to the $[-\pi/2, \pi/2]$ angular interval. 
Therefore, after receive combining of the $N'$ available scans $\eta_1, ..., \eta_{N'}$ as in \eqref{eq:RxBeamformer}, the available scan vector can be written as 
\begin{equation}
\overline{\mathbf{l}}_\text{(NA)} =  
\left[L_N\left(\eta_1 \right), ...,
L_N\left(\eta_{N'} \right) \right]
+ \mathbf{n} \; ,
\end{equation}
where the variance of each \gls{AWGN} sample in the row vector $\mathbf{n}$ is
\begin{equation}
\sigma^2 = \frac{1}{N^2}\sum_{j=0}^{N-1} \sigma_n^2 = \frac{1}{N}\sigma_n^2 \; , \forall i \;.
\label{eq:NoisePower}
\end{equation}
Hereafter, for ease of notation, we drop the number of sum co-array elements from the ideal angular response and received noisy response, leading to $L(\eta)$ and $R(\eta)$, respectively.

We place point scatterers in the environment such that their path losses and reflection coefficients are of unitary gain. 
Note that the estimation of few impulsive scatterers in the \gls{LAD} (or angular) domain, with their complex coefficients, coincides with channel estimation of narrowband mmWave and higher carrier frequency wireless channels, see for instance~\cite{alkhateeb2013hybrid}. The extension to broadband signals, thus requiring to estimate also the delays of the scatterers, will be subject to future work.
Our findings could be extended to more general cases of imaging, apart from the fact that performance, like object re\-so\-lu\-tion, is more difficult to measure. Assuming unitary power transmission, one could assess the performance impact with an equivalent \gls{SNR}
\begin{equation}
\gamma = \sigma^{-2} = N\sigma_n^{-2} \;.
\label{eq:SNR}
\end{equation}
Note that the factor $N$ corresponds to beamforming gain of the receive array. The $Q$ targets are placed randomly at angles in the $[-\theta^{\text{max}}, \theta^{\text{max}}]$ interval. Due to the typical $120$ degrees sector considered in deployments, we default to $\theta^\text{max} = 60$ degrees.
In some results, targets move at a constant speed of $\nu$ m/s, with a uniform random orientation with respect to the array, generating Doppler shifts due to acquisitions taken at different times.

Angles are scanned by steering both transmit and receive beamforming weights at the same angle, that changes at each acquisition. Given the $2N{-}1$ elements of the sum co-array, a minimum of $2N{-}1$ angles must be scanned. By default, the scanned angles are sampled uniformly in the \gls{LAD}, but we also evaluate the performance of uniform sampling in the angular domain. The angular sweeping is executed sequentially from the lowest \gls{LAD} value to the highest, with scan period $P_S$.
For the single target scenario, the target's \gls{LAD} location is estimated by up-sampling the angular response to 512 points and determining the location of its absolute value maximum. The estimate is further refined by determining the maximum's location of the parabola passing through the first determined maximum and its two neighbor points.

The investigated algorithms for up-sampling the \gls{LAD} response are
\begin{itemize}
\item The \gls{AR} proposal of Theorem \ref{theo:AR}, with $N'=2N-1$ \gls{LAD} samples. 
\item 
A CUBIC Hermite spline, which is a local interpolation using third-degree ($P{=}3$) polynomials that forces continuity on the interpolation's derivatives~\cite{zill2020advanced}, using $N'=2N-1$ \gls{LAD} samples. We chose this baseline due to its comparable computational complexity and better performance compared to other orders.
\item The ``\gls{SARA} - Red'' (reduced), where we instead used only $N'=N$ samples.
\end{itemize}
We also compare the \gls{LAD} estimation performance with the MUSIC algorithm~\cite{schmidt1986multiple}. The vector $\hat{\mathbf{a}}$ containing the complex signal at each of the $2N{-}1$ elements of the sum co-array can be estimated with the following \gls{MMSE} estimate~\cite{spagnolini2018statistical} as follows
\if\single1
\begin{subequations}\label{eq:MusicInversionGroup}
\begin{align}
\overline{\mathbf{l}}_\text{(NA)} &  
\overset{\eqref{eq:RxBeamformer}}{=}
{\underbrace{
\left[ a_0 \cdots a_{2N-1} \right]
}_\mathbf{a}}
{\underbrace{
\begin{bmatrix}
e^{-j 2 \pi x'_0\eta_1} & \cdots & e^{-j 2 \pi x'_0\eta_{2N-1}} \\ 
\vdots & \ddots & \vdots \\ 
e^{-j 2 \pi x'_{2N-1}\eta_1} & \cdots & e^{-j 2 \pi x'_{2N-1}\eta_{2N-1}}
\end{bmatrix}}_\mathbf{S}}
  + \mathbf{n}
\Rightarrow
\\
\hat{\mathbf{a}}&=  \overline{\mathbf{l}}_\text{(NA)} \mathbf{S}^H \left( \mathbf{S}\mathbf{S}^H + \sigma_n^2 \mathbf{I}_{2N-1}\right)^{-1} \;.
\label{eq:MusicAntennaSignal}
\end{align}
\end{subequations}
\else
\begin{subequations}\label{eq:MusicInversionGroup}
\begin{align}
\overline{\mathbf{l}}_\text{(NA)} &  
\overset{\eqref{eq:RxBeamformer}}{=}
{\underbrace{
\left[ a_0 \cdots a_{2N-1} \right]
}_\mathbf{a}}
{\underbrace{
\setlength\arraycolsep{2pt}
\begin{bmatrix}
e^{-j 2 \pi x'_0\eta_1} & {\cdots} & e^{-j 2 \pi x'_0\eta_{2N-1}} \\ 
\vdots & {\ddots} & \vdots \\ 
e^{-j 2 \pi x'_{2N-1}\eta_1} & {\cdots} & e^{-j 2 \pi x'_{2N-1}\eta_{2N-1}}
\end{bmatrix}}_\mathbf{S}}
\notag
\\
&+ \mathbf{n}
\Rightarrow 
\\
\hat{\mathbf{a}}&=  \overline{\mathbf{l}}_\text{(NA)} \mathbf{S}^H \left( \mathbf{S}\mathbf{S}^H + \sigma_n^2 \mathbf{I}_{2N-1}\right)^{-1} \;.
\label{eq:MusicAntennaSignal}
\end{align}
\end{subequations}
\fi
Note that the steering matrix $\mathbf{S}$ is a full rank square matrix with equal eigenvalues, when $2N-1$ scans in $\overline{\mathbf{l}}_\text{(NA)}$ are taken uniformly in \gls{LAD}. The multiple target search with MUSIC is performed as in Section III-B/C of~\cite{henninger2022computationally}. 

For the single target simulations, a single angle is estimated. On the other hand, in experiments with multiple targets, we estimate the number of scatterers $Q$ with the \gls{MDL} criterion~\cite{rissanen1978modeling} of the computed eigenvalues during MUSIC.


The complexity of CUBIC interpolation is $\mathcal{O}(NU)$ if the upsampling factor is much greater than the reconstruction order squared, i.e. $U \gg P^2$, otherwise $\mathcal{O}(NP^2)$~\cite{toraichi1987computational}. The computational effort of MUSIC is dominated by the evaluation of the MUSIC spectrum, for which evaluating a single point exhibits cubic complexity with respect to $2N{-}1$. However, to not miss any targets, the sampling granularity of the spectrum should be based on the resolution, requiring at least $2N{-}1$ evaluations, and thus resulting in $\mathcal{O}(N^4)$.
Accordingly, the asymptotic complexity of the considered algorithms can be found in Table~\ref{tab:ComplexityTheoretical}. One should note, however, that in our experiments we have observed almost an order of magnitude reduction in execution time of the \gls{AR} for all the considered values of $N$ and $U$ compared to CUBIC. On the other hand, the MUSIC super-resolution baseline show a larger scaling of the asymptotic behavior and a several orders of magnitude higher execution time in our numerical studies.


\begin{table*}[ht]%
\centering
\caption{Asymptotic complexity of the considered algorithms}
\label{tab:ComplexityTheoretical}
\begin{tabular}{|c|c|c|c|}
\hline
\gls{AR} & \gls{LR} & CUBIC & MUSIC \\ \hline
$\mathcal{O}\left(NU\log_2(NU)\right)$ & $\mathcal{O}(N^2U)$ & 
$\text{max} \left(\mathcal{O}(NU),\mathcal{O}(NP^2) \right)$ &
$\mathcal{O}(N^4)$  \\
\hline
\end{tabular}
\end{table*}

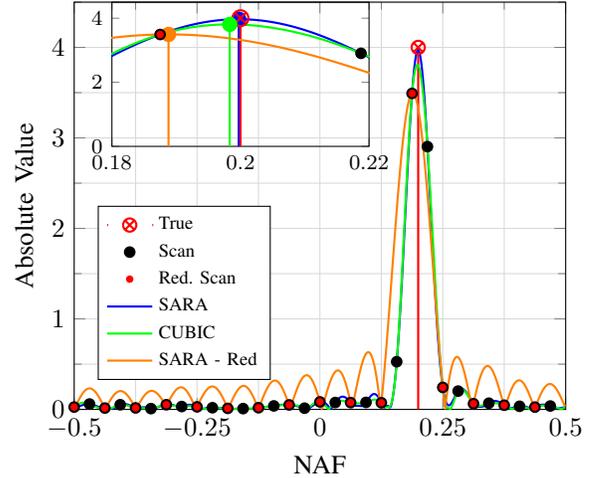
\begin{figure}[t!]
  \centering
  \input{Figures/ScanExample.tikz}
  \caption{Example of angular scan and reconstruction with a single target at \gls{LAD} $\ell = 0.2$ without noise, where  ``\gls{SARA} - Red'' uses  half of the angular scans compared to \gls{SARA} and CUBIC. } 
  \label{fig:ScanExample}
\end{figure}

Unless  mentioned otherwise, the default parameters used in every figure are recapped in Table~\ref{tab:SimDefaultParameters}. The scan period of $8.93$~$\mu s$ is due to the symbol length of $120$~kHz sub-carrier spacing, that is used at the $28$~GHz frequency bands~\cite{TS38.211}.
\begin{table}[ht]%
\centering
\caption{Default simulation parameters}
\label{tab:SimDefaultParameters}
\begin{tabular}{|c|c|}
\hline
Base station antennas $N$ & 16 vertically polarized
\\ \hline
Central frequency (wavelength $\lambda$) & $28$ GHz ($10.71$ mm)
\\ \hline
Antenna separation $d$ & $\lambda / 2$
\\ \hline
Scan period $P_S$ & $8.93$ $\mu$s
\\ \hline
Target speed $\nu$ & 0 m/s
\\ \hline
Maximum target's angle $\theta^{\text{max}}$ & $\pi/3$ rad. ($60$ deg.)
\\ \hline
\end{tabular}
\end{table}
An example of angular scan and reconstruction from a noise-less acquisition of a single target at \gls{LAD} $\ell = 0.2$ can be found in Fig.~\ref{fig:ScanExample}, where we plot the absolute values of the scan with points and the reconstruction curves, highlighting their maximum points.
For this figure, we chose a number of scans equal to $2N$ for \gls{SARA} and CUBIC, to allow angular samples of \gls{SARA} - Red (which only uses $N$ scans) to be in the same position as the other two alternatives.
Note how the \gls{SARA} (blue) maximum is on top of the true target's (red) \gls{LAD} position and amplitude, whereas the CUBIC reconstruction in green introduces some distortion. Finally, we have ``\gls{SARA} - Red'' using  only half of the angular scans that reconstructs the \gls{LAD} response with Dirichlet kernels with larger main lobe $B_N^{-1}$ half width - instead of $B_{2N-1}^{-1}$ - and order $N$. As shown earlier, this introduces aliasing in the \gls{AAL}. Therefore, the reconstructed response of ``\gls{SARA} - Red'' has a maximum with some displacement both in \gls{LAD} and amplitude with respect to the true target.

\subsection{Reconstruction results - Single target}\label{subsec:SimSingleTarget}
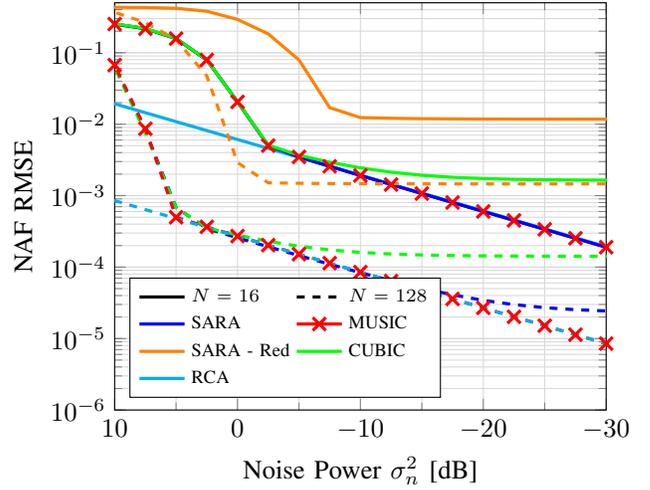
\begin{figure}
  \centering
  \input{Figures/RMSE_Variable_Antennas.tikz}
  \caption{\gls{RMSE} of single target \gls{LAD} estimation using \glspl{ULA} with variable number of 
  antennas. }

  \label{fig:RMSE_Variable_Antennas}
\end{figure}

In this subsection, we analyze the performance of angular estimation when a single target is present in the environment. This  allows to assess the capability of different algorithms without interference  by additional targets, being impacted only by the \gls{AWGN} and the algorithm itself.
\paragraph{\gls{LAD} \gls{RMSE} with variable $N$}
The \gls{RMSE} of a single target's \gls{LAD} estimate is plotted in Fig.~\ref{fig:RMSE_Variable_Antennas} with $N=16,128$. Note that the noise power axis is plotted in reverse order, corresponding to having plots with increasing \gls{SNR} on the x axis. The other parameters are according to Table \ref{tab:SimDefaultParameters}.
Given the \gls{LAD} period of $2d/\lambda = 1$, the \gls{LAD} errors are wrapped modulo $1$ operations, determining the \gls{LAD} \gls{RMSE} as follows
\begin{equation}
\text{\gls{LAD} \gls{RMSE}} = \sqrt{ E_j \left[ \left| \, \text{mod}_1 
\left( \eta_j - \hat{\eta}_j \right)  \right|^2 \right]} \; ,
\end{equation}
where $\eta_j, \hat{\eta}_j$ are the true and estimated \gls{LAD} in the $j$-th experiment, respectively. 
Given that the signal model at each antenna~\eqref{eq:ArrayLADResponse} coincides with a complex sinusoid, the problem of \gls{LAD} estimation can be seen as frequency estimation, given $N$ observations of a flat noisy channel. Therefore, assuming here to have access to the signal at each antenna element, which is possible only with digital beamforming capabilities, we can write the \gls{CRLB}, see equation~(26) of~\cite{baronkin2001cramer}.
\begin{equation}
\text{\gls{CRLB}} = \frac{1}{2 \pi \sigma_n}\sqrt{\frac{6}{N(N^2-1)}} \;.
\label{eq:CRLB}
\end{equation}
In contrast to typical angle of arrival bounds, \gls{CRLB} does not depend on the incoming angle, given the \gls{LAD} transformation representing the Fourier domain of the \gls{ULA}'s antenna locations.

Among the sampling and reconstruction algorithms, \gls{SARA} outperforms both CUBIC and ``\gls{SARA} - Red''. The latter is clearly suboptimal and will not be further evaluated. It exhibits a noise floor and is unable to capture the true characteristic of the \gls{LAD} response in its reconstruction. 
Even if the \gls{CRLB}~\eqref{eq:CRLB} is only an approximation of the lower bound on the performance of fully digital systems, we can note how \gls{SARA} approaches \gls{CRLB} curves at high \gls{SNR}, exhibiting a linear behavior with $\sigma_n^{-1}$. This is a further validation point for \gls{SARA}, which allows to achieve the same \gls{LAD} \gls{RMSE} of the \gls{CRLB}, as if digital beamforming were possible. However, this behavior starts to decline at extremely low \gls{RMSE}, due to \gls{LAD} quantization errors present in the up-sampled reconstruction with 512 points, instead of an ideal continuous function. 

{As a super-resolution method, MUSIC can deliver similar performance as \gls{SARA} at low \gls{SNR} and approach the \gls{CRLB} at high \gls{SNR}}. 
We can notice similar behavior with $N=16$ and $N=128$, where having more antennas simply allows to reduce the \gls{RMSE}.
From the discussion of this paragraph, in what follows, we choose CUBIC and MUSIC as our most interesting baselines. The first compares \gls{SARA} against another reconstruction algorithm, the latter against a super-resolution method.


%
%

\begin{figure}
  \centering
  \input{Figures/RMSE_Peak.tikz}
  \caption{\gls{RMSE} of single target peak estimation with variable number of antennas. 
  } 
  \label{fig:RMSE_Peak}
\end{figure}
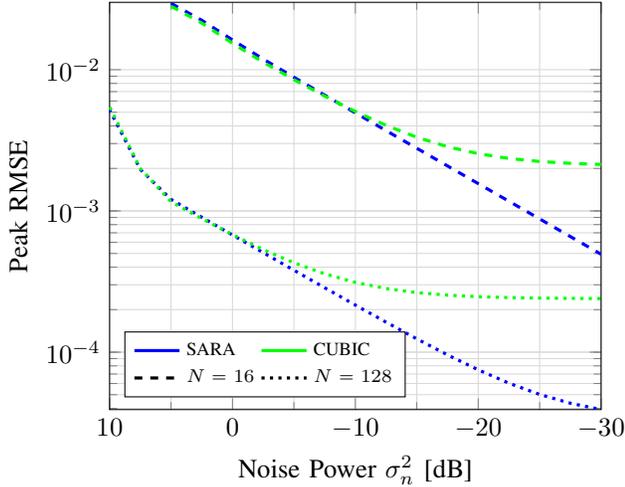

\paragraph{Target's peak amplitude \gls{RMSE} with variable $N$}
The \gls{RMSE} of the peak estimation error is plotted in Fig.~\ref{fig:RMSE_Peak} for the considered reconstruction algorithms, i.e. \gls{SARA} and CUBIC, and it is defined as
\begin{equation}
\text{Peak RMSE} = \sqrt{ E_j \left[ \left|  
 L(\eta_j) - R \left(\hat{\eta}_j \right)  \right|^2 \right]} \; ,
\end{equation}
where $R \left(\hat{\eta}_j \right)$ is the amplitude of the up-sampled reconstructed response, with the additional quadratic interpolation, as described in Subsection \ref{subsec:simScenario}. 
\gls{SARA} gains in high \gls{SNR} regimes, due to the absence of error floors, apart from the ones given by the reconstruction quantization, occurring at much lower Peak \gls{RMSE} values. Note that the noise power at which the two curves detach increases with the number of antennas.

\paragraph{The effect of user speed $\nu$}\label{par:SimUserSpeed}

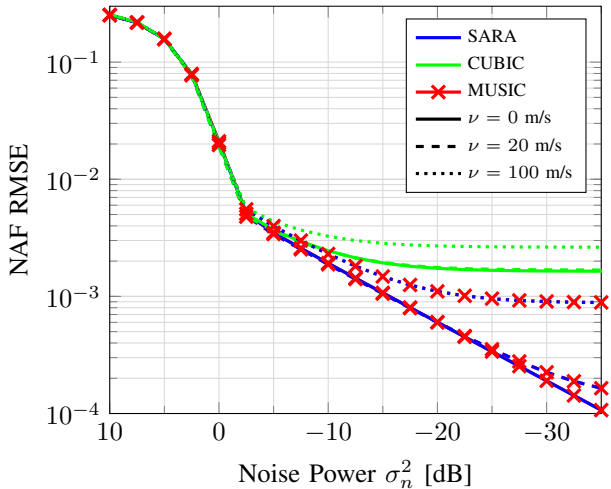
\begin{figure}
  \centering
\input{Figures/RMSE_Variable_Speed_MaxAngle60.tikz}
  \caption{\gls{RMSE} of single target \gls{LAD} estimation with target speed $\nu= \{0, 20, 100\}$ m/s.} 
  \label{fig:RMSE_Variable_Speed_MaxAngle60}
\end{figure}

We analyze in Fig.~\ref{fig:RMSE_Variable_Speed_MaxAngle60} the sensitivity with respect to target's speed $\nu$.
Note that the higher the speed values, the worse the performance degradation. Also in this case, the performances of \gls{SARA} and MUSIC overlap, while CUBIC exhibits a much higher error floor.


\paragraph{The effect of different angular sampling}
In this paragraph, we investigate the \gls{LAD} estimation losses due to sub-optimal uniform angular sampling, as done in~\cite{landmann2004efficient} - not in \gls{LAD} - in the following set
\begin{equation}
\mathcal{D}_{2N-1} = \left \{ \theta = \frac{n\pi}{N} : n \in \mathcal{N}_{2N-1} \right \} \;.
\end{equation}
The reconstruction is done working in the angular axes, leveraging Theorem \ref{theo:AR} (not in \gls{LAD} as stated in the theorem) for \gls{SARA} and simple cubic interpolation for the CUBIC case, whereas the signal at each antenna element for MUSIC is still estimated with~\eqref{eq:MusicAntennaSignal}.
We compare the ``Angle'' option versus the optimal uniform \gls{LAD} sampling of angles from the set $\mathcal{L}_{2N -1}$, according to \eqref{eq:AnglesToScanOnlyReceiver}, and reconstruction done in the \gls{LAD} domain.

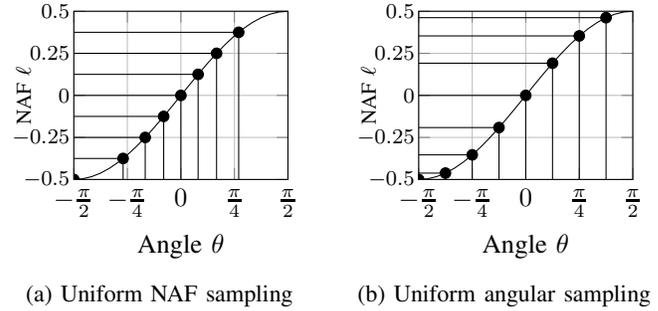
\begin{figure} 
\if\single1
        \centering
        \begin{subfigure}{0.45\textwidth}
            \input{Figures/UniformNAF_SingleColumn.tikz}
            \caption{Uniform \gls{LAD} sampling} 
            \label{subfig:UniformLADSampling}
        \end{subfigure}
        \begin{subfigure}{0.45\textwidth}
            \centering
            \input{Figures/UniformDeg_SingleColumn.tikz}
            \caption{Uniform angular sampling} 
            \label{subfig:UniformDegSampling}
        \end{subfigure}
\else
    \centering
  \subfloat[Uniform \gls{LAD} sampling\label{subfig:UniformLADSampling}]{%
  \input{Figures/UniformNAF.tikz}
       }
    \hfill
  \subfloat[Uniform angular sampling\label{subfig:UniformDegSampling}]{%
  \input{Figures/UniformDeg.tikz}
       }
\fi

  \caption{Example of uniform sampling of \gls{LAD} (a) and angular (b) domain with 8 samples. The plotted curve is the transformation between \gls{LAD} and angular domain $\ell = \frac{d}{\lambda}\sin(\theta) = \sin(\theta)/2$.}
  \label{fig:SamplingExperiments} 
\end{figure}

\begin{figure}
  \centering
  \input{Figures/RMSE_Variable_AngularSampling.tikz}
  \caption{\gls{RMSE} of single target \gls{LAD} estimation with uniform sampling of \gls{LAD} and angles.} 
  \label{fig:RMSE_Variable_AngularSampling}
\end{figure}
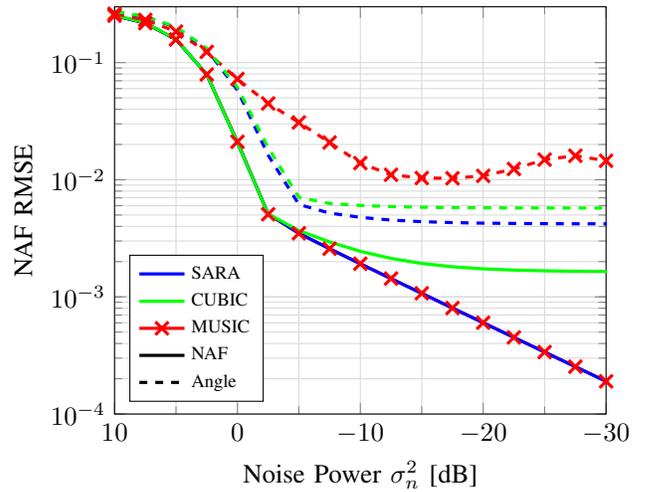

Fig.~\ref{fig:SamplingExperiments} shows an example of uniform \gls{LAD} and angular sampling with $8$ samples. Note that \gls{LAD} sampling focuses more on angles close to zero (array's boresight), allowing to achieve higher resolution at these angles. This is in line with the well-known property of \glspl{ULA} to have stronger resolution capabilities at boresight.
In Fig.~\ref{fig:RMSE_Variable_AngularSampling} we plot the \gls{LAD} estimation \gls{RMSE} for the two angular sampling criteria defined earlier.
One can observe the error floors of dashed ``Angle'' curves, implying that sampling angles uniformly in the angular domain leads to performance losses. The performance gap is more remarkable with MUSIC compared to the other two considered algorithms, due to the difficulty of equalizing the steering vector matrix $\mathbf{S}$ in~\eqref{eq:MusicInversionGroup} if the $2N-1$ sampled angles are not uniform in \gls{LAD}.

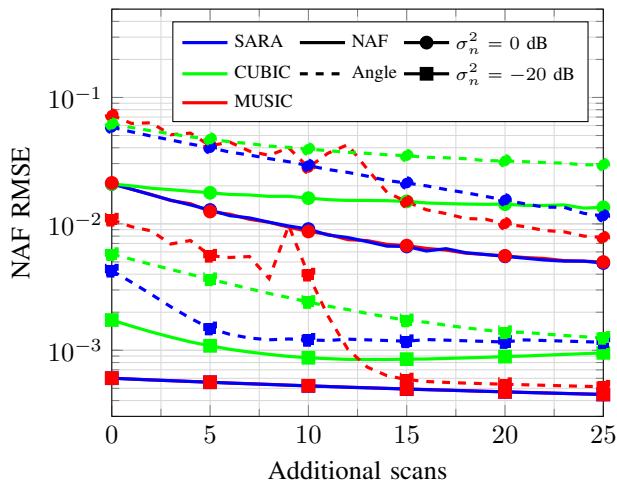
\begin{figure}[h]
  \centering
\input{Figures/RMSE_OverProv_Variable_AngularSampling_NoisePower.tikz}
  \caption{\gls{RMSE} of single target \gls{LAD} estimation versus the number of additional scans over the minimum number of $2N -1 = 31$. Curves are with uniform sampling of \gls{LAD}  and angles, and with different noise power $\sigma_n^2=\{0, -20\}$ dB.} 
  \label{fig:RMSE_OverProv_VariableAngularSampling_NoisePower}
\end{figure}

\paragraph{The impact of additional scans}
Additional angular acquisitions on top of the minimum $2N-1$ can be considered to reduce the impact of poor angular sampling choices, see the Angle curves of Fig.~\ref{fig:RMSE_Variable_AngularSampling}. For example, in Fig.~\ref{fig:RMSE_OverProv_VariableAngularSampling_NoisePower} the \gls{LAD} estimation \gls{RMSE} is plotted versus the number of scans used minus $2N-1$. The curves are for different noise powers 0 dB and $-20$ dB and angular sampling techniques, i.e. the \gls{LAD} and Angle uniform sampling introduced in the previous paragraph.

As one could anticipate, the \gls{RMSE} diminishes with additional scans. However, the performance of the optimal ``\gls{SARA} - \gls{LAD}'' (and the overlapping ``MUSIC - \gls{LAD}'') improves only slightly since the signal was not experiencing any distortion in the reconstruction according to Theorem \ref{theo:AR}. Therefore, the additional angular samples just increase
the \gls{SNR}, as described in Subsection \ref{subsec:DifferentNumberScans}. On the other hand, the gains in all the other options are larger, thanks to the increased number of samples taken close to boresight, i.e. $\ell = 0$, that is where \gls{LAD} samples the angular domain more frequently compared to Angle. Focusing on the low noise case, i.e., $\sigma_n=-20$ dB, we see that ``\gls{SARA} - Angle'' and ``CUBIC - Angle'' still exhibit an error floor. On the other hand, after ca. 15 additional scans (50\% more) the \gls{MMSE} equalizer in~\eqref{eq:MusicAntennaSignal} seems to finally be able to properly equalize $\mathbf{S}$, allowing the MUSIC Angle curve to approach the overlapping ``\gls{SARA} - \gls{LAD}'' and ``MUSIC - \gls{LAD}'' curves.


%


%% file: Figures/ScanExample.tikz.tex
\tikzset{mark size=2.5}    
    \begin{tikzpicture}
	\begin{axis}[
    		height = 7cm,
    		xlabel={\gls{LAD}},
    		ylabel={Absolute Value},
    		ymin=0,
		ymax=4.5,
		xmin=-0.5,
		xmax=0.5,
		minor tick num=1,
		yminorticks = true,
		enlargelimits = false,
		grid = both,
    		grid style={solid, black!15},
    		legend columns = 1,
    		legend style={at={(0.4,0.5)}, font=\scriptsize},
		xtick = {-0.5, -0.25, 0, 0.25, 0.5},
        tick label style={/pgf/number format/fixed},
		legend cell align={left},
		every axis plot/.append style={thick}
		]

		\addplot[ycomb, mark=otimes, color=red] plot table[x index=0, y expr=\thisrowno{1}/4.0]{Data/ExportedTargetTrue.txt};
		
		\addplot[only marks, mark=*, mark size = 1.8, color=black] plot table[x index=0, y expr=\thisrowno{1}/4.0] {Data/ExportedScan.txt};
		\addplot[only marks, mark=*, mark size = 1, color=red, each nth point=2] plot table[x index=0, y expr=\thisrowno{1}/4.0] {Data/ExportedScan.txt};
		
		\addplot[color=blue] plot table[x index=0, y expr=\thisrowno{1}/4.0] {Data/ExportedCurves.txt};
		\addplot[color=green] plot table[x index=0, y expr=\thisrowno{2}/4.0] {Data/ExportedCurves.txt};
		\addplot[color=orange] plot table[x index=0, y expr=\thisrowno{3}/4.0] {Data/ExportedCurves.txt};
		
	\legend{True, Scan, Red. Scan, \gls{SARA}, CUBIC, \gls{SARA} - Red}
   	
   	\end{axis} 
   	
   	\begin{axis}[ymin=0,
		ymax=4.5,
		xmin=0.18,
		xmax=0.22,
		footnotesize,
		height=3.5cm, 
		width=5cm,
		xtick = {0.18, 0.2, 0.22},
		at={(0.5cm, 3.5cm)},
		every axis plot/.append style={thick} ]

		\addplot[ycomb, mark=*, color=blue] plot table[x index=0, y expr=\thisrowno{1}/4.0]{Data/ExportedTargetFFT.txt};
		
		\addplot[ycomb, mark=otimes, color=red, mark size = 3] plot table[x index=0, y expr=\thisrowno{1}/4.0]{Data/ExportedTargetTrue.txt};
		
		\addplot[ycomb, mark=*, color=green] plot table[x index=0, y expr=\thisrowno{1}/4.0]{Data/ExportedTargetCubic.txt};
		
		\addplot[ycomb, mark=*, color=orange] plot table[x index=0, y expr=\thisrowno{1}/4.0]{Data/ExportedTargetReduced.txt};
		
		\addplot[only marks, mark=*, mark size = 1.8, color=black] plot table[x index=0, y expr=\thisrowno{1}/4.0] {Data/ExportedScan.txt};
		\addplot[only marks, mark=*, mark size = 1, color=red, each nth point=2] plot table[x index=0, y expr=\thisrowno{1}/4.0] {Data/ExportedScan.txt};
		
		\addplot[color=blue] plot table[x index=0, y expr=\thisrowno{1}/4.0] {Data/ExportedCurves.txt};
		\addplot[color=green] plot table[x index=0, y expr=\thisrowno{2}/4.0] {Data/ExportedCurves.txt};
		\addplot[color=orange] plot table[x index=0, y expr=\thisrowno{3}/4.0] {Data/ExportedCurves.txt};
    \end{axis}   

\end{tikzpicture}

%
%
%
%
%
%

%% file: Figures/RMSE_Variable_Antennas.tikz.tex
\tikzset{mark size=3.5}    
    \begin{tikzpicture}
	\begin{semilogyaxis}[
    		height = 7cm,
    		xlabel={Noise Power $\sigma_n^2$ [dB]},
    		ylabel={\gls{LAD} \gls{RMSE}},
    	ymin=1e-6,
		ymax=5e-1,
		x dir=reverse,
		xmin=-30.0,
		xmax=10.0,
		minor tick num=1,
		yminorticks = true,
		enlargelimits = false,
		legend pos=south west, 
		grid = both,
    		grid style={solid, black!15},
    		legend columns = 2,
    		legend style={font=\scriptsize},
        tick label style={/pgf/number format/fixed},
		legend cell align={left},
		every axis plot/.append style={very thick}
		]

%

\def\antennas{16}
	
		\addplot[color=red, solid, forget plot] plot table[x index=0, y index=5] {Data/NAF_RMSE_ANTENNAS_\antennas.txt};		\addplot[color=cyan, solid, forget plot] plot table[x index=0, y index=1] {Data/NAF_CRLB_ANTENNAS_\antennas.txt};
		\addplot[color=blue, solid, forget plot] plot table[x index=0, y index=1] {Data/NAF_RMSE_ANTENNAS_\antennas.txt};
		\addplot[color=orange, solid, forget plot] plot table[x index=0, y index=2] {Data/NAF_RMSE_ANTENNAS_\antennas_Reduced.txt};
		\addplot[color=green, solid, forget plot] plot table[x index=0, y index=3] {Data/NAF_RMSE_ANTENNAS_\antennas.txt};
		\addplot[color=red, solid, forget plot, only marks, mark=x] plot table[x index=0, y index=5] {Data/NAF_RMSE_ANTENNAS_\antennas.txt};

\def\antennas{128}
	
		\addplot[color=blue, dashed, forget plot] plot table[x index=0, y index=1] {Data/NAF_RMSE_ANTENNAS_\antennas.txt};
		\addplot[color=orange, dashed, forget plot] plot table[x index=0, y index=2] {Data/NAF_RMSE_ANTENNAS_\antennas_Reduced.txt};
		\addplot[color=green, dashed, forget plot] plot table[x index=0, y index=3] {Data/NAF_RMSE_ANTENNAS_\antennas.txt};
		\addplot[color=red, dashed, forget plot] plot table[x index=0, y index=5] {Data/NAF_RMSE_ANTENNAS_\antennas.txt};	
		\addplot[color=red, only marks, forget plot, mark=x] plot table[x index=0, y index=5] {Data/NAF_RMSE_ANTENNAS_\antennas.txt};	
		\addplot[color=cyan, dashed, forget plot] plot table[x index=0, y index=1] {Data/NAF_CRLB_ANTENNAS_\antennas.txt};

\def\legy{2.0}

		\addplot[color=black, solid, draw=none] coordinates {(0, \legy)};
		\addplot[color=black, dashed, draw=none] coordinates {(0, \legy)};
		\addplot[color=blue, solid, draw=none] coordinates {(0, \legy)};
		\addplot[color=red, solid, draw=none, mark=x] coordinates {(0, \legy)};
		\addplot[color=orange, solid, draw=none] coordinates {(0, \legy)};
		\addplot[color=green, solid, draw=none] coordinates {(0, \legy)};
		\addplot[color=cyan, solid, draw=none] coordinates {(0, \legy)};

	\legend{$N=16$, $N=128$, \gls{SARA}, MUSIC, \gls{SARA} - Red, CUBIC, \gls{CRLB}}
   	
   	\end{semilogyaxis} 

\end{tikzpicture}

%
%
%
%
%
%

%% file: Figures/RMSE_Peak.tikz.tex
\tikzset{mark size=2.5}    
    \begin{tikzpicture}
	\begin{semilogyaxis}[
    		height = 7cm,
    		xlabel={Noise Power $\sigma_n^2$ [dB]},
    		ylabel={Peak \gls{RMSE}},
		ymax=3e-2,
		x dir=reverse,
		xmin=-30.0,
		xmax=10.0,
		minor tick num=1,
		yminorticks = true,
		enlargelimits = false,
		legend pos=south west,
		grid = both,
    		grid style={solid, black!15},
    		legend columns = 2,
    		legend style={font=\scriptsize},
        tick label style={/pgf/number format/fixed},
		legend cell align={left},
		every axis plot/.append style={very thick}
		]

%

\def\antennas{16}
	
		\addplot[color=blue, dashed, forget plot] plot table[x index=0, y index=1] {Data/Peak_RMSE_ANTENNAS_\antennas.txt};
		\addplot[color=green, dashed, forget plot] plot table[x index=0, y index=2] {Data/Peak_RMSE_ANTENNAS_\antennas.txt};

	

\def\antennas{128}
	
		\addplot[color=blue, dotted, forget plot] plot table[x index=0, y index=1] {Data/Peak_RMSE_ANTENNAS_\antennas_Reduced.txt};
		\addplot[color=green, dotted, forget plot] plot table[x index=0, y index=2] {Data/Peak_RMSE_ANTENNAS_\antennas_Reduced.txt};

		\addplot[color=blue, solid, draw=none] coordinates {(0, 0.1)};
		\addplot[color=green, solid, draw=none] coordinates {(0, 0.1)};
		\addplot[color=black, dashed, draw=none] coordinates {(0, 0.1)};
		\addplot[color=black, dotted, draw=none] coordinates {(0, 0.1)};
		
	\legend{\gls{SARA}, CUBIC, $N=16$, $N=128$}
   	
   	\end{semilogyaxis} 

\end{tikzpicture}

%
%
%
%
%
%

%% file: Figures/RMSE_Variable_Speed_MaxAngle60.tikz.tex
\tikzset{mark size=3.5}    
    \begin{tikzpicture}
	\begin{semilogyaxis}[
    		height = 7cm,
    		xlabel={Noise Power $\sigma_n^2$ [dB]},
    		ylabel={\gls{LAD} \gls{RMSE}},
    	ymin=1e-4,
		ymax=3e-1,
		x dir=reverse,
		xmin=-35.0,
		xmax=10.0,
		minor tick num=1,
		yminorticks = true,
		enlargelimits = false,
		legend pos=north east,
		grid = both,
    		grid style={solid, black!15},
    		legend columns = 1,
    		legend style={font=\scriptsize},
        tick label style={/pgf/number format/fixed},
		legend cell align={left},
		every axis plot/.append style={very thick}
		]

\def\maxangle{"60.0"}
\def\speed{"0.0"}
		
		\addplot[color=red, solid, forget plot] plot table[x index=0, y index=3] {Data/NAF_RMSE_SPEED_SPEED_\speed_MANGLE_\maxangle.txt};
		\addplot[color=blue, solid, forget plot] plot table[x index=0, y index=1] {Data/NAF_RMSE_SPEED_SPEED_\speed_MANGLE_\maxangle.txt};
		\addplot[color=green, solid, forget plot] plot table[x index=0, y index=2] {Data/NAF_RMSE_SPEED_SPEED_\speed_MANGLE_\maxangle.txt};	
		\addplot[color=red, solid, forget plot, only marks,  mark=x] plot table[x index=0, y index=3] {Data/NAF_RMSE_SPEED_SPEED_\speed_MANGLE_\maxangle.txt};
			
\def\speed{"20.0"}
	
		\addplot[color=red, dashed, forget plot] plot table[x index=0, y index=3] {Data/NAF_RMSE_SPEED_SPEED_\speed_MANGLE_\maxangle.txt};
		\addplot[color=blue, dashed, forget plot] plot table[x index=0, y index=1] {Data/NAF_RMSE_SPEED_SPEED_\speed_MANGLE_\maxangle.txt};
		\addplot[color=green, dashed, forget plot] plot table[x index=0, y index=2] {Data/NAF_RMSE_SPEED_SPEED_\speed_MANGLE_\maxangle.txt};
		\addplot[color=red, solid, forget plot, only marks,  mark=x] plot table[x index=0, y index=3] {Data/NAF_RMSE_SPEED_SPEED_\speed_MANGLE_\maxangle.txt};
		
\def\speed{"100.0"}
	
		\addplot[color=red, dotted, forget plot] plot table[x index=0, y index=3] {Data/NAF_RMSE_SPEED_SPEED_\speed_MANGLE_\maxangle.txt};
		\addplot[color=blue, dotted, forget plot] plot table[x index=0, y index=1] {Data/NAF_RMSE_SPEED_SPEED_\speed_MANGLE_\maxangle.txt};
		\addplot[color=green, dotted, forget plot] plot table[x index=0, y index=2] {Data/NAF_RMSE_SPEED_SPEED_\speed_MANGLE_\maxangle.txt};
		\addplot[color=red, solid, forget plot, only marks,  mark=x] plot table[x index=0, y index=3] {Data/NAF_RMSE_SPEED_SPEED_\speed_MANGLE_\maxangle.txt};

\def\legy{2.0}
		
		\addplot[color=blue, solid, draw=none] coordinates {(0, \legy)};
		\addplot[color=green, solid, draw=none] coordinates {(0, \legy)};
	    \addplot[color=red, solid, mark=x, draw=none] coordinates {(0, \legy)};
		\addplot[color=black, solid, draw=none] coordinates {(0, \legy)};
		\addplot[color=black, dashed, draw=none] coordinates {(0, \legy)};
		\addplot[color=black, dotted, draw=none] coordinates {(0, \legy)};
		
	\legend{\gls{SARA}, CUBIC, MUSIC, $\nu=0$ m/s, $\nu=20$ m/s, $\nu=100$ m/s}
   	
   	\end{semilogyaxis} 

\end{tikzpicture}

%
%
%
%
%
%

%% file: Figures/UniformNAF_SingleColumn.tikz.tex
\begin{tikzpicture}
    
      \begin{axis}[width=1.0\textwidth,
      			   xlabel={Angle $\theta$},
      			   ylabel={\gls{LAD} $\ell$},
      			   xmin=-0.5,
      			   xmax=0.5,
      			   ymin=-0.5,
      			   ymax=0.5,
      			   xtick={-.5,-.25,...,.5},
      			   xticklabels={$-\frac{\pi}{2}$,$-\frac{\pi}{4}$,$0$,$\frac{\pi}{4}$,$\frac{\pi}{2}$},
      			   ytick={-0.5,-0.25,0,0.25, 0.5},
      			   ylabel shift = -14 pt,
      			   ylabel style={font=\normalsize},
      			   yticklabel style={font=\normalsize,xshift=2pt},
      			   xticklabel style={font=\normalsize},
      			   grid=both
      			   ]
		\addplot[domain=-.5:.5, samples, color=black]{0.5*sin(deg(\x * pi))};

\foreach \naf in {-0.5,-0.375,...,0.375}
{
	\pgfmathsetmacro{\deg}{rad(asin(\naf * 2)) / pi}
    \edef\temp{
    	\noexpand\draw[-, color=black] (\deg,-0.5) -- (\deg,\naf) -- (-0.5,\naf) ;
    	\noexpand\filldraw[color=black] (\deg,\naf) circle (2pt);}
    \temp
}		

      \end{axis}
    \end{tikzpicture}

%% file: Figures/UniformDeg_SingleColumn.tikz.tex
\begin{tikzpicture}

    
      \begin{axis}[width=1.0\textwidth,
      			   xlabel={Angle $\theta$},
      			   ylabel={\gls{LAD} $\ell$},
      			   xmin=-0.5,
      			   xmax=0.5,
      			   ymin=-0.5,
      			   ymax=0.5,
      			   xtick={-.5,-.25,...,.5},
      			   xticklabels={$-\frac{\pi}{2}$,$-\frac{\pi}{4}$,$0$,$\frac{\pi}{4}$,$\frac{\pi}{2}$},
      			   ytick={-0.5,-0.25,0,0.25, 0.5},
      			   ylabel shift = -14 pt,
      			   ylabel style={font=\normalsize},
      			   yticklabel style={font=\normalsize,xshift=2pt},
      			   xticklabel style={font=\normalsize},
      			   grid=both
      			   ]
		\addplot[domain=-.5:.5, samples, color=black]{0.5*sin(deg(\x * pi))};


\foreach \x in {-.5,-.375,...,.375}
{
	\pgfmathsetmacro{\naf}{0.5 * sin(deg(\x *pi))}
    \edef\temp{
    	\noexpand\draw[-, color=black] (\x,-0.5) -- (\x,\naf) -- (-0.5,\naf);
    	\noexpand\filldraw[color=black] (\x,\naf) circle (2pt);}
    \temp
}	

      \end{axis}
    \end{tikzpicture}

%% file: Figures/UniformNAF.tikz.tex
\begin{tikzpicture}
    
      \begin{axis}[width=0.5\linewidth,
      			   xlabel={Angle $\theta$},
      			   ylabel={\gls{LAD} $\ell$},
      			   xmin=-0.5,
      			   xmax=0.5,
      			   ymin=-0.5,
      			   ymax=0.5,
      			   xtick={-.5,-.25,...,.5},
      			   xticklabels={$-\frac{\pi}{2}$,$-\frac{\pi}{4}$,$0$,$\frac{\pi}{4}$,$\frac{\pi}{2}$},
      			   ytick={-0.5,-0.25,0,0.25, 0.5},
      			   ylabel shift = -14 pt,
      			   ylabel style={font=\footnotesize},
      			   yticklabel style={font=\footnotesize,xshift=2pt},
      			   xticklabel style={font=\normalsize},
      			   grid=both
      			   ]
		\addplot[domain=-.5:.5, samples, color=black]{0.5*sin(deg(\x * pi))};

\foreach \naf in {-0.5,-0.375,...,0.375}
{
	\pgfmathsetmacro{\deg}{rad(asin(\naf * 2)) / pi}
    \edef\temp{
    	\noexpand\draw[-, color=black] (\deg,-0.5) -- (\deg,\naf) -- (-0.5,\naf) ;
    	\noexpand\filldraw[color=black] (\deg,\naf) circle (2pt);}
    \temp
}		

      \end{axis}
    \end{tikzpicture}

%% file: Figures/UniformDeg.tikz.tex
\begin{tikzpicture}

    
      \begin{axis}[width=0.5\linewidth,
      			   xlabel={Angle $\theta$},
      			   ylabel={\gls{LAD} $\ell$},
      			   xmin=-0.5,
      			   xmax=0.5,
      			   ymin=-0.5,
      			   ymax=0.5,
      			   xtick={-.5,-.25,...,.5},
      			   xticklabels={$-\frac{\pi}{2}$,$-\frac{\pi}{4}$,$0$,$\frac{\pi}{4}$,$\frac{\pi}{2}$},
      			   ytick={-0.5,-0.25,0,0.25, 0.5},
      			   ylabel shift = -14 pt,
      			   ylabel style={font=\footnotesize},
      			   yticklabel style={font=\footnotesize,xshift=2pt},
      			   xticklabel style={font=\normalsize},
      			   grid=both
      			   ]
		\addplot[domain=-.5:.5, samples, color=black]{0.5*sin(deg(\x * pi))};


\foreach \x in {-.5,-.375,...,.375}
{
	\pgfmathsetmacro{\naf}{0.5 * sin(deg(\x *pi))}
    \edef\temp{
    	\noexpand\draw[-, color=black] (\x,-0.5) -- (\x,\naf) -- (-0.5,\naf);
    	\noexpand\filldraw[color=black] (\x,\naf) circle (2pt);}
    \temp
}	

      \end{axis}
    \end{tikzpicture}

%% file: Figures/RMSE_Variable_AngularSampling.tikz.tex
\tikzset{mark size=3.5}    
    \begin{tikzpicture}
	\begin{semilogyaxis}[
    		height = 7cm,
    		xlabel={Noise Power $\sigma_n^2$ [dB]},
    		ylabel={\gls{LAD} \gls{RMSE}},
    	ymin=1e-4,
		ymax=3e-1,
		x dir=reverse,
		xmin=-30.0,
		xmax=10.0,
		minor tick num=1,
		yminorticks = true,
		enlargelimits = false,
		legend pos=south west,
		grid = both,
    		grid style={solid, black!15},
    		legend columns = 1,
    		legend style={font=\scriptsize},
        tick label style={/pgf/number format/fixed},
		legend cell align={left},
		every axis plot/.append style={very thick}
		]

\def\sampling{"NAF"}
	
		\addplot[color=red, solid, forget plot] plot table[x index=0, y index=3] {Data/NAF_RMSE_SAMPLING_DOMAIN_\sampling.txt};
		\addplot[color=blue, solid, forget plot] plot table[x index=0, y index=1] {Data/NAF_RMSE_SAMPLING_DOMAIN_\sampling.txt};
		\addplot[color=green, solid, forget plot] plot table[x index=0, y index=2] {Data/NAF_RMSE_SAMPLING_DOMAIN_\sampling.txt};
		\addplot[color=red, solid, forget plot, only marks,  mark=x] plot table[x index=0, y index=3] {Data/NAF_RMSE_SAMPLING_DOMAIN_\sampling.txt};

\def\sampling{"Deg"}
	
		\addplot[color=red, dashed, forget plot] plot table[x index=0, y index=3] {Data/NAF_RMSE_SAMPLING_DOMAIN_\sampling.txt};
		\addplot[color=blue, dashed, forget plot] plot table[x index=0, y index=1] {Data/NAF_RMSE_SAMPLING_DOMAIN_\sampling.txt};
		\addplot[color=green, dashed, forget plot] plot table[x index=0, y index=2] {Data/NAF_RMSE_SAMPLING_DOMAIN_\sampling.txt};
		\addplot[color=red, solid, forget plot, only marks,  mark=x] plot table[x index=0, y index=3] {Data/NAF_RMSE_SAMPLING_DOMAIN_\sampling.txt};

\def\legy{2.0}
		\addplot[color=blue, solid, draw=none] coordinates {(0, \legy)};
		\addplot[color=green, solid, draw=none] coordinates {(0, \legy)};
		\addplot[color=red, solid, draw=none, mark=x] coordinates {(0, \legy)};
		\addplot[color=black, solid, draw=none] coordinates {(0, \legy)};
		\addplot[color=black, dashed, draw=none] coordinates {(0, \legy)};
		
	\legend{\gls{SARA}, CUBIC, MUSIC, \gls{LAD}, Angle}
   	
   	\end{semilogyaxis} 

\end{tikzpicture}

%
%
%
%
%
%

%% file: Figures/RMSE_OverProv_Variable_AngularSampling_NoisePower.tikz.tex
\tikzset{mark size=2.0}    
    \begin{tikzpicture}
	\begin{semilogyaxis}[
    		height = 7cm,
    		xlabel={Additional scans},
    		ylabel={\gls{LAD} \gls{RMSE}},
    		ymin=0,
		ymax=0.5,
		ymin=3e-4,
		xmin=0,
		xmax=25,
		minor tick num=1,
		yminorticks = true,
		enlargelimits = false,
		legend pos=north east,
		grid = both,
    		grid style={solid, black!15},
    		legend columns = 3,
    		legend style={font=\scriptsize},
        tick label style={/pgf/number format/fixed},
		legend cell align={left},
		every axis plot/.append style={very thick},
		mark repeat={5}
		]

\def\noise{"0.0"}
\def\sampling{"NAF"}
	
		\addplot[color=red, solid, forget plot] plot table[x index=0, y index=3] {Data/NAF_RMSE_OVERPROV_NOISE_\noise_DOMAIN_\sampling.txt};
		\addplot[color=blue, solid, forget plot, mark=*] plot table[x index=0, y index=1] {Data/NAF_RMSE_OVERPROV_NOISE_\noise_DOMAIN_\sampling.txt};
		\addplot[color=green, solid, forget plot, mark=*] plot table[x index=0, y index=2] {Data/NAF_RMSE_OVERPROV_NOISE_\noise_DOMAIN_\sampling.txt};
		\addplot[color=red, solid, forget plot, only marks, mark=*] plot table[x index=0, y index=3] {Data/NAF_RMSE_OVERPROV_NOISE_\noise_DOMAIN_\sampling.txt};

\def\sampling{"Deg"}
	
		\addplot[color=red, dashed, forget plot, mark=*] plot table[x index=0, y index=3] {Data/NAF_RMSE_OVERPROV_NOISE_\noise_DOMAIN_\sampling.txt};
		\addplot[color=blue, dashed, forget plot, mark=*] plot table[x index=0, y index=1] {Data/NAF_RMSE_OVERPROV_NOISE_\noise_DOMAIN_\sampling.txt};
		\addplot[color=green, dashed, forget plot, mark=*] plot table[x index=0, y index=2] {Data/NAF_RMSE_OVERPROV_NOISE_\noise_DOMAIN_\sampling.txt};
		
\def\noise{"-20.0"}
\def\sampling{"NAF"}
	
		\addplot[color=red, solid, forget plot] plot table[x index=0, y index=3] {Data/NAF_RMSE_OVERPROV_NOISE_\noise_DOMAIN_\sampling.txt};
		\addplot[color=blue, solid, forget plot, mark=square*] plot table[x index=0, y index=1] {Data/NAF_RMSE_OVERPROV_NOISE_\noise_DOMAIN_\sampling.txt};
		\addplot[color=green, solid, forget plot, mark=square*] plot table[x index=0, y index=2] {Data/NAF_RMSE_OVERPROV_NOISE_\noise_DOMAIN_\sampling.txt};
		\addplot[color=red, solid, forget plot, mark=square*, only marks] plot table[x index=0, y index=3] {Data/NAF_RMSE_OVERPROV_NOISE_\noise_DOMAIN_\sampling.txt};

\def\sampling{"Deg"}
	
		\addplot[color=red, dashed, forget plot, mark=square*] plot table[x index=0, y index=3] {Data/NAF_RMSE_OVERPROV_NOISE_\noise_DOMAIN_\sampling.txt};
		\addplot[color=blue, dashed, forget plot, mark=square*] plot table[x index=0, y index=1] {Data/NAF_RMSE_OVERPROV_NOISE_\noise_DOMAIN_\sampling.txt};
		\addplot[color=green, dashed, forget plot, mark=square*] plot table[x index=0, y index=2] {Data/NAF_RMSE_OVERPROV_NOISE_\noise_DOMAIN_\sampling.txt};

		\addplot[color=blue, solid, draw=none] coordinates {(0, 1)};
		\addplot[color=black, solid, draw=none] coordinates {(0, 1)};
		\addplot[color=black, solid, mark=*, draw=none] coordinates {(0, 1)};
		\addplot[color=green, solid, draw=none] coordinates {(0, 1)};
		\addplot[color=black, dashed, draw=none] coordinates {(0, 1)};
		\addplot[color=black, solid, mark=square*, draw=none] coordinates {(0, 1)};
		\addplot[color=red, solid, draw=none] coordinates {(0, 1)};
		
	\legend{\gls{SARA}, \gls{LAD}, $\sigma_n^2 = 0$ dB, CUBIC, Angle, $\sigma_n^2 = -20$ dB, MUSIC}
   	
   	\end{semilogyaxis} 

\end{tikzpicture}

%
%
%
%
%
%

%% file: Content/MultipleTargets.tex
\subsection{Reconstruction results - Multiple targets}

\if\single1
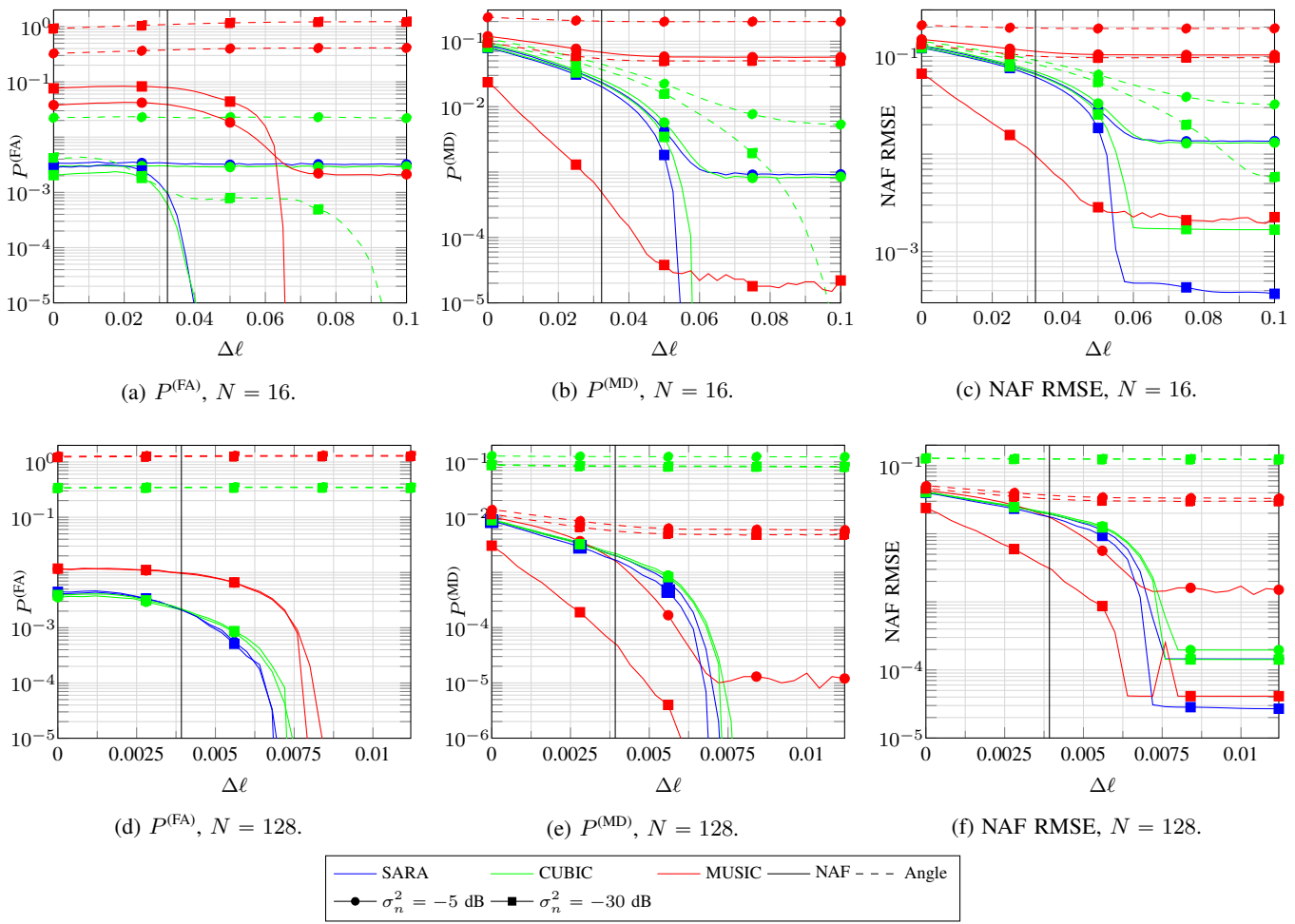
\begin{figure*}[ht]
        \centering
        \begin{subfigure}[t]{0.325\textwidth}
            \centering
			\input{Figures/FA_Rate_NAF_Sweep_16.tikz}
            \caption{$P^\text{(FA)}$, $N=16$.} 
            \label{fig:FA_Rate_NAF_Sweep_16}
        \end{subfigure}
        \hfil
        \begin{subfigure}[t]{0.325\textwidth}
            \centering
			\input{Figures/Missed_Det_NAF_Sweep_16.tikz}
		    \vspace{-0.46cm}
            \caption{$P^\text{(MD)}$, $N=16$.} 
            \label{fig:Missed_Det_NAF_Sweep_16}
        \end{subfigure}
        \hfil
        \begin{subfigure}[t]{0.325\textwidth}
            \centering
			\input{Figures/RMSE_NAF_Sweep_16.tikz}
            \caption{\gls{LAD} \gls{RMSE}, $N=16$.}  
            \label{fig:RMSE_NAF_Sweep_16}
        \end{subfigure}
        \vskip\baselineskip
        \vspace{-0.5cm}
        \begin{subfigure}[t]{0.325\textwidth}
            \centering
			\input{Figures/FA_Rate_NAF_Sweep_128.tikz}
			\vspace{-0.46cm}
            \caption{$P^\text{(FA)}$, $N=128$.} 
            \label{fig:FA_Rate_NAF_Sweep_128}
            \vspace{0.3cm}
        \end{subfigure}
        \hfil
        \begin{subfigure}[t]{0.325\textwidth}
            \centering
			\input{Figures/Missed_Det_NAF_Sweep_128.tikz}
            \caption{$P^\text{(MD)}$, $N=128$.} 
            \label{fig:Missed_Det_NAF_Sweep_128}
        \end{subfigure}
        \hfil
        \begin{subfigure}[t]{0.325\textwidth}
            \centering
			\input{Figures/RMSE_NAF_Sweep_128.tikz}
            \caption{\gls{LAD} \gls{RMSE}, $N=128$.}  
            \label{fig:RMSE_NAF_Sweep_128}
        \end{subfigure}
        \centerline{\ref{named}}
\else
\begin{figure*}[ht]
        \centering
        \begin{subfigure}[t]{0.325\textwidth}
            \centering
			\input{Figures/FA_Rate_NAF_Sweep_16.tikz}
            \caption{$P^\text{(FA)}$, $N=16$.} 
            \label{fig:FA_Rate_NAF_Sweep_16}
        \end{subfigure}
        \hfil
        \begin{subfigure}[t]{0.325\textwidth}
            \centering
			\input{Figures/Missed_Det_NAF_Sweep_16.tikz}
		    \vspace{-0.45cm}
            \caption{$P^\text{(MD)}$, $N=16$.} 
            \label{fig:Missed_Det_NAF_Sweep_16}
        \end{subfigure}
        \hfil
        \begin{subfigure}[t]{0.325\textwidth}
            \centering
			\input{Figures/RMSE_NAF_Sweep_16.tikz}
            \caption{\gls{LAD} \gls{RMSE}, $N=16$.}  
            \label{fig:RMSE_NAF_Sweep_16}
        \end{subfigure}
        \vskip\baselineskip
        \begin{subfigure}[t]{0.325\textwidth}
            \centering
			\input{Figures/FA_Rate_NAF_Sweep_128.tikz}
			\vspace{-0.45cm}
            \caption{$P^\text{(FA)}$, $N=128$.} 
            \label{fig:FA_Rate_NAF_Sweep_128}
            \vspace{0.2cm}
        \end{subfigure}
        \hfil
        \begin{subfigure}[t]{0.325\textwidth}
            \centering
			\input{Figures/Missed_Det_NAF_Sweep_128.tikz}
            \caption{$P^\text{(MD)}$, $N=128$.} 
            \label{fig:Missed_Det_NAF_Sweep_128}
        \end{subfigure}
        \hfil
        \begin{subfigure}[t]{0.325\textwidth}
            \centering
			\input{Figures/RMSE_NAF_Sweep_128.tikz}
            \caption{\gls{LAD} \gls{RMSE}, $N=128$.}  
            \label{fig:RMSE_NAF_Sweep_128}
        \end{subfigure}
        \centerline{\ref{named}}
\fi
        \caption{Probability of false alarm, probability of missed detection, and \gls{LAD} \gls{RMSE} of a scenario with three targets and increasing minimum \gls{LAD} spacing between them. All \gls{SARA} curves are with uniform sampling in the \gls{LAD} domain (solid), while for CUBIC and MUSIC also uniform sampling of the angles is considered (dashed). The different noise powers $\sigma_n^2=\{-5, -30\}$~dB are marked with circles, and squares, respectively. The vertical lines indicate the Rayleigh criterion for resolution at $\Delta \ell=1/(2N-1)$.} 
\label{fig:NAF_Sweep}
\end{figure*}
In this subsection, we consider experiments with $Q=3$ targets placed at random angles, but spaced by at least $\Delta \ell$ in the \gls{LAD} domain. Therefore, for the reconstruction techniques (\gls{SARA} and CUBIC), we assume that a target is present if the angular response's magnitude exceeds the \gls{CFAR} threshold given by
\begin{equation}
\zeta = \sqrt{-\sigma^2 \ln(1-(1-P^\text{(FA)})^{\frac{1}{2N -1}})} \; ,
\label{eq:CFAR}
\end{equation}
where $P^\text{(FA)}$ is the desired false alarm probability, defined as the likelihood of erroneously detecting a target in the response ~\cite{richards2014principles}.
\if\onappendix1
For the derivation of \eqref{eq:CFAR}, please refer to Appendix~\ref{app:CFARDerivation}.
\else
For the derivation of \eqref{eq:CFAR}, please refer to Appendix~D of the supplementary material, where we compute $\zeta$ such that the probability of a noise sample's amplitude to exceed it is the desired $P^\text{(FA)}$.
\fi
In our study, we chose $P^\text{(FA)} = 10^{-3}$. Note that we also consider a detected peak a false alarm if it is further away from a true target than the 
Dirichlet kernel's main lobe half width of $1/(2N-1)$, corresponding to the Rayleigh criterion for the resolution of two targets~\cite{johnson1992array}.
The reconstructed angular response is iteratively scanned for the strongest peak. We coherently remove the $i$-th iteration's target contribution to get the angular response for the next iteration as
\begin{equation}
R_{i+1}(l) = R_{i}(l) - R_i\left(\hat{\eta}_i \right) \cdot  \left( D_{N} \left( \ell - \hat{\eta}_i \right) \right)^2 \;,
\label{eq:PeakRemoval}
\end{equation}
where $\hat{\eta}_i$ is the \gls{LAD} estimate of the $i$-th iteration's target. Notice the squaring of $D_N(l)$, which is due to the multiplying effects of transmitter and receiver (see Subsection \ref{subsec:DirectionalTxRx}). However, canceling peaks generated by multiple interfering targets may create false residual peaks at iterations $i > 1$. Their detection is prevented by rejecting peak estimates if there was no peak in the original $R_1(l), l \in [\hat{\eta}_i - 1/(2N-1), \hat{\eta}_i + 1/(2N-1)]$.
Finally, sidelobes and inter-target interference can cause additional detections, especially in low-noise scenarios where $\zeta^{(\text{CFAR})}$ is low. To account for this, the threshold is updated after the first peak's detection as 
\begin{equation}
\zeta' = \max{\Big\{\zeta , \; \kappa \cdot S_{N} \cdot R_1\left(\hat{\eta}_1 \right)\Big\}} \;,
\label{eq:CFARAdaption}
\end{equation}
where $S_N$ is the first sidelobe's relative amplitude of the used Dirichlet kernel \cite{johnson1992array}, and $\kappa$ a scaling factor chosen based on experiments and tuned such that the desired $P^\text{(FA)}$ can be approximately attained. We found $\kappa = 6$ to be a suitable value in our scenario. On the other hand, for MUSIC, the peak search routine returns the $\hat{Q}$ strongest peaks, where $\hat{Q}$ is estimated with the \gls{MDL} criterion~\cite{rissanen1978modeling}.


Fig.~\ref{fig:NAF_Sweep} shows probability of false alarm $P^\text{(FA)}$, probability of missed detection $P^\text{(MD)}$, defined as the percentage of undetected targets, and \gls{LAD} \gls{RMSE} with $N=16$ and $N=128$.
We compare our proposal \gls{SARA} with uniform \gls {LAD} sampling 
against CUBIC and MUSIC, with uniform sampling either in \gls{LAD} or  angular domain (``Angle''),
respectively. Each curve is evaluated at noise powers $\sigma_n^2$ of $-5$~dB and $-30$~dB. 
The vertical curves at $\Delta \ell=1/(2N-1)$ representing the Rayleigh criterion for the resolution of two targets are also added.



Observing the $P^\text{(FA)}$ curves in the first column of Fig. \ref{fig:NAF_Sweep}, it can be seen that, due to the imperfect model order estimation with \gls{MDL}, using MUSIC leads to a higher number of false detections than the reconstruction techniques ``\gls{SARA} - \gls{LAD}'' and ``CUBIC - \gls{LAD}'', which exhibit comparable false alarm rates. In low-noise scenarios, using uniform \gls{LAD} sampling generally prevents false alarms for high enough $\Delta \ell$, whereas uniform sampling in the angular domain can lead to uncontrollable false alarm rates. The $P^\text{(MD)}$ curves in Figs.~\ref{fig:Missed_Det_NAF_Sweep_16} and \ref{fig:Missed_Det_NAF_Sweep_128} show that MUSIC, thanks to its super-resolution properties, offers advantages in terms of resolving closely spaced targets in high \gls{SNR} regimes. Between the reconstruction techniques, ``\gls{SARA} - \gls{LAD}'' offers minor, but consistent resolution gains compared to ``CUBIC - \gls{LAD}''. Also here, the benefits of sampling the \gls{LAD} domain uniformly are visible, as ``CUBIC - Angle'' and ``MUSIC - Angle'' exhibit significantly higher missed detection probabilities than their respective \gls{LAD} counterparts. Finally, the \gls{LAD} \gls{RMSE} curves (rightmost column of Fig. \ref{fig:NAF_Sweep}) display similar patterns compared to the previously analyzed $P^\text{(MD)}$ results. Nonetheless, ``\gls{SARA} - \gls{LAD}'' achieves the lowest \gls{RMSE} error floor for high target separation, even though giving up some resolution capability with respect to MUSIC in low-noise regimes at small $\Delta \ell$. It is further important to highlight that high-noise ($\sigma_n^2 = -5$~dB) impacts MUSIC the most, especially for $N=16$, where its capability diminishes significantly, while both reconstruction techniques ``\gls{SARA} - \gls{LAD}'' and ``CUBIC - \gls{LAD}'' still perform reasonably well.

%% file: Figures/FA_Rate_NAF_Sweep_16.tikz.tex
\if\single1
\def\scale{.635}
\else
\def\scale{.72}
\fi

\tikzset{mark size=2.5}    
    \begin{tikzpicture}
	\begin{semilogyaxis}[
    		xlabel={$\Delta \ell$},
    		ylabel={$P^\text{(FA)}$},
    	ylabel shift = -14 pt,
      	ylabel style={font=\footnotesize,at={(axis description cs:.-0.05,.5)},rotate=0,anchor=south},
      	ytick={0.00001, 0.00002, 0.00003, 0.00004, 0.00005, 0.00006, 0.00007, 0.00008, 0.00009, 0.0001, 0.0002, 0.0003, 0.0004, 0.0005, 0.0006, 0.0007, 0.0008, 0.0009, 0.001, 0.002, 0.003, 0.004, 0.005, 0.006, 0.007, 0.008, 0.009, 0.01, 0.02, 0.03, 0.04, 0.05, 0.06, 0.07, 0.08, 0.09, 0.1, 0.2, 0.3, 0.4, 0.5, 0.6, 0.7, 0.8, 0.9, 1, 2, 3, 4, 5, 6, 7, 8, 9, 10},
      	yticklabels={$10^{-5}$, , , , , , , , ,$10^{-4}$, , , , , , , , ,$10^{-3}$, , , , , , , , , , , , , , , , , ,$10^{-1}$ , , , , , , , , , $10^{0}$},
        yticklabel style={font=\footnotesize,xshift=3pt},
        xticklabel style={font=\footnotesize},
      	xlabel style={font=\footnotesize},
    	ymin=1e-5,
		ymax=2,
		minor tick num=1,
		yminorticks = true,
		enlargelimits = false,
		legend pos = north east,
		vasymptote=0.032258,
		grid = both,
		grid style={solid, black!15},
		legend columns = 6,
		transpose legend,
		legend style={font=\scriptsize},
        tick label style={/pgf/number format/fixed},
        ticklabel style={font=\footnotesize},
		legend cell align={left},
		every axis plot/.append style={very thick},
		mark repeat=10,
		scale = \scale
		]

\def\antennas{"16"}

\def\noise{"-5.0"}

\def\k{"6.0"}

		\addplot[thin, color=blue, solid, mark=*, mark options={scale=0.75}, forget plot] plot table[x index=0, y index=1] {Data/multTargets/FA_RATE_MIN_NAF_SPACING_NOISE_\noise_ANTENNAS_\antennas_SAMPLING_DOMAIN_NAF_KAPPA_\k.txt};
		\addplot[thin, color=green, solid, mark=*, mark options={scale=0.75}, forget plot] plot table[x index=0, y index=2] {Data/multTargets/FA_RATE_MIN_NAF_SPACING_NOISE_\noise_ANTENNAS_\antennas_SAMPLING_DOMAIN_NAF_KAPPA_\k.txt};
		\addplot[thin, color=green, dashed, mark=*, mark options={scale=0.75}, forget plot] plot table[x index=0, y index=2] {Data/multTargets/FA_RATE_MIN_NAF_SPACING_NOISE_\noise_ANTENNAS_\antennas_SAMPLING_DOMAIN_DEG_KAPPA_\k.txt};
		\addplot[thin, color=red, solid, mark=*, mark options={scale=0.75}, forget plot] plot table[x index=0, y index=3] {Data/multTargets/FA_RATE_MIN_NAF_SPACING_NOISE_\noise_ANTENNAS_\antennas_SAMPLING_DOMAIN_NAF_KAPPA_\k.txt};
		\addplot[thin, color=red, dashed, mark=*, mark options={scale=0.75}, forget plot] plot table[x index=0, y index=3] {Data/multTargets/FA_RATE_MIN_NAF_SPACING_NOISE_\noise_ANTENNAS_\antennas_SAMPLING_DOMAIN_DEG_KAPPA_\k.txt};


\def\noise{"-30.0"}

		\addplot[thin, color=blue, solid, mark=square*, mark options={scale=0.75}, mark phase = 6, forget plot] plot table[x index=0, y index=1] {Data/multTargets/FA_RATE_MIN_NAF_SPACING_NOISE_\noise_ANTENNAS_\antennas_SAMPLING_DOMAIN_NAF_KAPPA_\k.txt};
		\addplot[thin, color=green, solid, mark=square*, mark options={scale=0.75}, mark phase = 7, forget plot] plot table[x index=0, y index=2] {Data/multTargets/FA_RATE_MIN_NAF_SPACING_NOISE_\noise_ANTENNAS_\antennas_SAMPLING_DOMAIN_NAF_KAPPA_\k.txt};
		\addplot[thin, color=green, dashed, mark=square*, mark options={scale=0.75}, mark phase = 8, forget plot] plot table[x index=0, y index=2] {Data/multTargets/FA_RATE_MIN_NAF_SPACING_NOISE_\noise_ANTENNAS_\antennas_SAMPLING_DOMAIN_DEG_KAPPA_\k.txt};
		\addplot[thin, color=red, solid, mark=square*, mark options={scale=0.75}, mark phase = 7, forget plot] plot table[x index=0, y index=3] {Data/multTargets/FA_RATE_MIN_NAF_SPACING_NOISE_\noise_ANTENNAS_\antennas_SAMPLING_DOMAIN_NAF_KAPPA_\k.txt};
		\addplot[thin, color=red, dashed, mark=square*, mark options={scale=0.75}, forget plot] plot table[x index=0, y index=3] {Data/multTargets/FA_RATE_MIN_NAF_SPACING_NOISE_\noise_ANTENNAS_\antennas_SAMPLING_DOMAIN_DEG_KAPPA_\k.txt};


		
\addplot[thin, color=black, dotted, draw=none] coordinates {(0, 1)}; 
\addplot[thin, color=black, solid, draw=none] coordinates {(0, 1)}; 
   	
\end{semilogyaxis} 

\end{tikzpicture}
	

%% file: Figures/Missed_Det_NAF_Sweep_16.tikz.tex
\if\single1
\def\scale{.635}
\else
\def\scale{.72}
\fi

\tikzset{mark size=2.5}    
    \begin{tikzpicture}
	\begin{semilogyaxis}[
    		xlabel={$\Delta \ell$},
    		ylabel={$P^\text{(MD)}$},
      	ylabel shift = -14 pt,
      	ylabel style={font=\footnotesize,at={(axis description cs:.-0.05,.5)},rotate=0,anchor=south},
      	ytick={0.00001, 0.00002, 0.00003, 0.00004, 0.00005, 0.00006, 0.00007, 0.00008, 0.00009, 0.0001, 0.0002, 0.0003, 0.0004, 0.0005, 0.0006, 0.0007, 0.0008, 0.0009, 0.001, 0.002, 0.003, 0.004, 0.005, 0.006, 0.007, 0.008, 0.009, 0.01, 0.02, 0.03, 0.04, 0.05, 0.06, 0.07, 0.08, 0.09, 0.1, 0.2, 0.3, 0.4},
      	yticklabels={$10^{-5}$, , , , , , , , , $10^{-4}$, , , , , , , , , , , , , , , , , ,$10^{-2}$, , , , , , , , , $10^{-1}$},
      	xlabel style={font=\footnotesize},
        yticklabel style={font=\footnotesize,xshift=3pt},
        xticklabel style={font=\footnotesize},
    	ymin=1e-5,
		ymax=3e-1,
		minor tick num=1,
		yminorticks = true,
		enlargelimits = false,
		legend pos = south west,
		vasymptote=0.032258,
		grid = both,
    		grid style={solid, black!15},
    		legend columns = 5,
    		legend style={font=\scriptsize},
        tick label style={/pgf/number format/fixed},
		legend cell align={left},
		every axis plot/.append style={very thick},
		mark repeat={10},
		legend to name = named,
		scale = \scale
		]

\def\antennas{"16"}
\def\kappa{"6.0"}
\def\noise{"-5.0"}
	
		\addplot[thin, color=blue, solid, mark=*, mark options={scale=0.75}, forget plot] plot table[x index=0, y index=1] {Data/multTargets/MISSED_DET_MIN_NAF_SPACING_NOISE_\noise_ANTENNAS_\antennas_SAMPLING_DOMAIN_NAF_KAPPA_\kappa.txt};
		\addplot[thin, color=green, solid, mark=*, mark options={scale=0.75}, forget plot] plot table[x index=0, y index=2] {Data/multTargets/MISSED_DET_MIN_NAF_SPACING_NOISE_\noise_ANTENNAS_\antennas_SAMPLING_DOMAIN_NAF_KAPPA_\kappa.txt};
		\addplot[thin, color=green, dashed, mark=*, mark options={scale=0.75}, forget plot] plot table[x index=0, y index=2] {Data/multTargets/MISSED_DET_MIN_NAF_SPACING_NOISE_\noise_ANTENNAS_\antennas_SAMPLING_DOMAIN_Deg_KAPPA_\kappa.txt};
		\addplot[thin, color=red, solid, mark=*, mark options={scale=0.75}, forget plot] plot table[x index=0, y index=3] {Data/multTargets/MISSED_DET_MIN_NAF_SPACING_NOISE_\noise_ANTENNAS_\antennas_SAMPLING_DOMAIN_NAF_KAPPA_\kappa.txt};
		\addplot[thin, color=red, dashed, mark=*, mark options={scale=0.75}, forget plot] plot table[x index=0, y index=3] {Data/multTargets/MISSED_DET_MIN_NAF_SPACING_NOISE_\noise_ANTENNAS_\antennas_SAMPLING_DOMAIN_DEG_KAPPA_\kappa.txt};

\def\noise{"-30.0"}

		\addplot[thin, color=blue, solid, mark=square*, mark options={scale=0.75}, forget plot, mark phase = 2] plot table[x index=0, y index=1] {Data/multTargets/MISSED_DET_MIN_NAF_SPACING_NOISE_\noise_ANTENNAS_\antennas_SAMPLING_DOMAIN_NAF_KAPPA_\kappa.txt};
		\addplot[thin, color=green, solid, mark=square*, mark options={scale=0.75}, forget plot, mark phase = 4] plot table[x index=0, y index=2] {Data/multTargets/MISSED_DET_MIN_NAF_SPACING_NOISE_\noise_ANTENNAS_\antennas_SAMPLING_DOMAIN_NAF_KAPPA_\kappa.txt};
		\addplot[thin, color=green, dashed, mark=square*, mark options={scale=0.75}, forget plot, mark phase = 9] plot table[x index=0, y index=2] {Data/multTargets/MISSED_DET_MIN_NAF_SPACING_NOISE_\noise_ANTENNAS_\antennas_SAMPLING_DOMAIN_Deg_KAPPA_\kappa.txt};
		\addplot[thin, color=red, solid, mark=square*, mark options={scale=0.75}, forget plot] plot table[x index=0, y index=3] {Data/multTargets/MISSED_DET_MIN_NAF_SPACING_NOISE_\noise_ANTENNAS_\antennas_SAMPLING_DOMAIN_NAF_KAPPA_\kappa.txt};
		\addplot[thin, color=red, dashed, mark=square*, mark options={scale=0.75}, forget plot] plot table[x index=0, y index=3] {Data/multTargets/MISSED_DET_MIN_NAF_SPACING_NOISE_\noise_ANTENNAS_\antennas_SAMPLING_DOMAIN_DEG_KAPPA_\kappa.txt};


		\addplot[thin, color=blue, solid, draw=none] coordinates {(0, 1)}; \addlegendentry{\gls{SARA}}
		\addplot[thin, color=green, solid, draw=none] coordinates {(0, 1)}; \addlegendentry{CUBIC}
		\addplot[thin, color=red, solid, draw=none] coordinates {(0, 1)}; \addlegendentry{MUSIC}
		\addplot[thin, color=black, solid, draw=none] coordinates {(0, 1)}; \addlegendentry{\gls{LAD}}
		\addplot[thin, color=black, dashed, draw=none] coordinates {(0, 1)}; \addlegendentry{Angle}
		\addplot[thin, color=black, solid, mark=*, mark options={scale=0.75}, draw=none] coordinates {(0, 1)};\addlegendentry{$\sigma_n^2 = -5$ dB}
		\addplot[thin, color=black, solid, mark=square*, mark options={scale=0.75}, draw=none] coordinates {(0, 1)};\addlegendentry{$\sigma_n^2 = -30$ dB}
   	
   	\end{semilogyaxis} 

\end{tikzpicture}

%% file: Figures/RMSE_NAF_Sweep_16.tikz.tex
\if\single1
\def\scale{.635}
\else
\def\scale{.72}
\fi

\tikzset{mark size=2.5}    
    \begin{tikzpicture}
	\begin{semilogyaxis}[
    		xlabel={$\Delta \ell$},
    		ylabel={\gls{LAD} \gls{RMSE}},
    	ylabel shift = -14 pt,
      	ylabel style={font=\footnotesize,at={(axis description cs:.-0.05,.5)},rotate=0,anchor=south},
      	ytick={0.0004, 0.0005, 0.0006, 0.0007, 0.0008, 0.0009, 0.001, 0.002, 0.003, 0.004, 0.005, 0.006, 0.007, 0.008, 0.009, 0.01, 0.02, 0.03, 0.04, 0.05, 0.06, 0.07, 0.08, 0.09, 0.1, 0.2, 0.3},
      	yticklabels={, , , , , , $10^{-3}$, , , , , , , , , , , , , , , , , , $10^{-1}$},
      	xlabel style={font=\footnotesize},
        yticklabel style={font=\footnotesize,xshift=3pt},
        xticklabel style={
        	/pgf/number format/precision=2,
  			/pgf/number format/fixed, 
  			font=\footnotesize},
    	ymin=3e-4,
		ymax=3e-1,
		minor tick num=1,
		yminorticks = true,
		enlargelimits = false,
		legend pos = south west,
		vasymptote=0.032258,
		grid = both,
    		grid style={solid, black!15},
    		legend columns = 6,
    		transpose legend,
    		legend style={font=\scriptsize},
        tick label style={/pgf/number format/fixed},
        ticklabel style={font=\footnotesize},
		legend cell align={left},
		every axis plot/.append style={very thick},
		mark repeat={10},
		scale = \scale
		]

\def\antennas{"16"}
\def\kappa{"6.0"}
\def\noise{"-5.0"}
	
		\addplot[thin, color=blue, solid, mark=*, mark options={scale=0.75}, forget plot] plot table[x index=0, y index=1] {Data/multTargets/NAF_RMSE_MIN_NAF_SPACING_\noise_ANTENNAS_\antennas_SAMPLING_DOMAIN_NAF_KAPPA_\kappa.txt};
		\addplot[thin, color=green, solid, mark=*, mark options={scale=0.75}, forget plot] plot table[x index=0, y index=2] {Data/multTargets/NAF_RMSE_MIN_NAF_SPACING_\noise_ANTENNAS_\antennas_SAMPLING_DOMAIN_NAF_KAPPA_\kappa.txt};
		\addplot[thin, color=green, dashed, mark=*, mark options={scale=0.75}, forget plot] plot table[x index=0, y index=2] {Data/multTargets/NAF_RMSE_MIN_NAF_SPACING_\noise_ANTENNAS_\antennas_SAMPLING_DOMAIN_Deg_KAPPA_\kappa.txt};
		\addplot[thin, color=red, solid, mark=*, mark options={scale=0.75}, forget plot] plot table[x index=0, y index=3] {Data/multTargets/NAF_RMSE_MIN_NAF_SPACING_\noise_ANTENNAS_\antennas_SAMPLING_DOMAIN_NAF_KAPPA_\kappa.txt};
		\addplot[thin, color=red, dashed, mark=*, mark options={scale=0.75}, forget plot] plot table[x index=0, y index=3] {Data/multTargets/NAF_RMSE_MIN_NAF_SPACING_\noise_ANTENNAS_\antennas_SAMPLING_DOMAIN_DEG_KAPPA_\kappa.txt};

\def\noise{"-30.0"}

		\addplot[thin, color=blue, solid, mark=square*, mark options={scale=0.75}, forget plot] plot table[x index=0, y index=1] {Data/multTargets/NAF_RMSE_MIN_NAF_SPACING_\noise_ANTENNAS_\antennas_SAMPLING_DOMAIN_NAF_KAPPA_\kappa.txt};
		\addplot[thin, color=green, solid, mark=square*, mark options={scale=0.75}, forget plot] plot table[x index=0, y index=2] {Data/multTargets/NAF_RMSE_MIN_NAF_SPACING_\noise_ANTENNAS_\antennas_SAMPLING_DOMAIN_NAF_KAPPA_\kappa.txt};
		\addplot[thin, color=green, dashed, mark=square*, mark options={scale=0.75}, forget plot] plot table[x index=0, y index=2] {Data/multTargets/NAF_RMSE_MIN_NAF_SPACING_\noise_ANTENNAS_\antennas_SAMPLING_DOMAIN_Deg_KAPPA_\kappa.txt};
		\addplot[thin, color=red, solid, mark=square*, mark options={scale=0.75}, forget plot] plot table[x index=0, y index=3] {Data/multTargets/NAF_RMSE_MIN_NAF_SPACING_\noise_ANTENNAS_\antennas_SAMPLING_DOMAIN_NAF_KAPPA_\kappa.txt};
		\addplot[thin, color=red, dashed, mark=square*, mark options={scale=0.75}, forget plot] plot table[x index=0, y index=3] {Data/multTargets/NAF_RMSE_MIN_NAF_SPACING_\noise_ANTENNAS_\antennas_SAMPLING_DOMAIN_DEG_KAPPA_\kappa.txt};

		\addplot[thin, color=blue, solid, draw=none] coordinates {(0, 1)}; \addlegendentry{\gls{SARA} \gls{LAD}}
		\addplot[thin, color=green, solid, draw=none] coordinates {(0, 1)}; \addlegendentry{CUBIC \gls{LAD}}
		\addplot[thin, color=red, solid, draw=none] coordinates {(0, 1)}; \addlegendentry{MUSIC \gls{LAD}}
		\addplot[thin, color=green, dashed, draw=none] coordinates {(0, 1)}; 
        \addlegendentry{CUBIC Angle}
		\addplot[thin, color=red, dashed, draw=none] coordinates {(0, 1)}; \addlegendentry{MUSIC Angle}
		\addplot[thin, color=black, solid, mark=*, mark options={scale=0.75}, draw=none] coordinates {(0, 1)};\addlegendentry{$\sigma_n^2 = -5$ dB}
		\addplot[color=black, solid, mark=square*, mark options={scale=0.75}, draw=none] coordinates {(0, 1)};\addlegendentry{$\sigma_n^2 = -30$ dB}
   	
\legend{}

\end{semilogyaxis} 

\end{tikzpicture}

		

%% file: Figures/FA_Rate_NAF_Sweep_128.tikz.tex
\if\single1
\def\scale{.635}
\else
\def\scale{.72}
\fi

\tikzset{mark size=2.5}    
    \begin{tikzpicture}
	\begin{semilogyaxis}[
    		xlabel={$\Delta \ell$},
    		ylabel={$P^\text{(FA)}$},
    	ylabel shift = -14 pt,
      	ylabel style={font=\footnotesize,at={(axis description cs:.-0.05,.5)},rotate=0,anchor=south},
      	ytick={0.00001, 0.00002, 0.00003, 0.00004, 0.00005, 0.00006, 0.00007, 0.00008, 0.00009, 0.0001, 0.0002, 0.0003, 0.0004, 0.0005, 0.0006, 0.0007, 0.0008, 0.0009, 0.001, 0.002, 0.003, 0.004, 0.005, 0.006, 0.007, 0.008, 0.009, 0.01, 0.02, 0.03, 0.04, 0.05, 0.06, 0.07, 0.08, 0.09, 0.1, 0.2, 0.3, 0.4, 0.5, 0.6, 0.7, 0.8, 0.9, 1, 2, 3, 4, 5, 6},
      	yticklabels={$10^{-5}$, , , , , , , , ,$10^{-4}$, , , , , , , , ,$10^{-3}$, , , , , , , , , , , , , , , , , ,$10^{-1}$, , , , , , , , , $10^{0}$},
        yticklabel style={font=\footnotesize,xshift=3pt},
        xticklabel style={
        	/pgf/number format/precision=4,
  			/pgf/number format/fixed, 
  			font=\footnotesize},
      	xlabel style={font=\footnotesize},
    	ymin=1e-5,
		ymax=2,
		xmin=0,
		xmax=0.0112,
		minor tick num=1,
		yminorticks = true,
		enlargelimits = false,
		legend pos = south west,
		scaled x ticks = false,
		vasymptote=0.003921,
		grid = both,
		grid style={solid, black!15},
		legend columns = 6,
		transpose legend,
        legend style={
            at={(0.41,0.15)},
            anchor=southwest, 
            font=\scriptsize},
		xtick = {0, 0.0025, 0.005, 0.0075, 0.01},
		xticklabels = {0, 0.0025, 0.005, 0.0075, 0.01},
        ticklabel style={font=\footnotesize},
        legend cell align={left},
		every axis plot/.append style={very thick},
		mark repeat={7},
		scale = \scale
		]

\def\antennas{"128"}
\def\k{"6.0"}
\def\noise{"-5.0"}
	
		\addplot[thin, color=blue, solid, mark=*, mark options={scale=0.75},mark phase = 4, forget plot] plot table[x index=0, y index=1] {Data/multTargets/FA_RATE_MIN_NAF_SPACING_NOISE_\noise_ANTENNAS_\antennas_SAMPLING_DOMAIN_NAF_KAPPA_\k.txt};
		\addplot[thin, color=green, solid, mark=*, mark options={scale=0.75}, mark phase = 5, forget plot] plot table[x index=0, y index=2] {Data/multTargets/FA_RATE_MIN_NAF_SPACING_NOISE_\noise_ANTENNAS_\antennas_SAMPLING_DOMAIN_NAF_KAPPA_\k.txt};
		\addplot[thin, color=green, dashed, mark=*, mark options={scale=0.75}, forget plot] plot table[x index=0, y index=2] {Data/multTargets/FA_RATE_MIN_NAF_SPACING_NOISE_\noise_ANTENNAS_\antennas_SAMPLING_DOMAIN_DEG_KAPPA_\k.txt};
		\addplot[thin, color=red, solid, mark=*, mark options={scale=0.75}, mark phase = 7, forget plot] plot table[x index=0, y index=3] {Data/multTargets/FA_RATE_MIN_NAF_SPACING_NOISE_\noise_ANTENNAS_\antennas_SAMPLING_DOMAIN_NAF_KAPPA_\k.txt};
		\addplot[thin, color=red, dashed, mark=*, mark options={scale=0.75}, forget plot] plot table[x index=0, y index=3] {Data/multTargets/FA_RATE_MIN_NAF_SPACING_NOISE_\noise_ANTENNAS_\antennas_SAMPLING_DOMAIN_DEG_KAPPA_\k.txt};

\def\noise{"-30.0"}

		\addplot[thin, color=blue, solid, mark=square*, mark options={scale=0.75}, mark phase = 4,forget plot] plot table[x index=0, y index=1] {Data/multTargets/FA_RATE_MIN_NAF_SPACING_NOISE_\noise_ANTENNAS_\antennas_SAMPLING_DOMAIN_NAF_KAPPA_\k.txt};
		\addplot[thin, color=green, solid, mark=square*, mark options={scale=0.75}, mark phase = 5, forget plot] plot table[x index=0, y index=2] {Data/multTargets/FA_RATE_MIN_NAF_SPACING_NOISE_\noise_ANTENNAS_\antennas_SAMPLING_DOMAIN_NAF_KAPPA_\k.txt};
		\addplot[thin, color=green, dashed, mark=square*, mark options={scale=0.75}, forget plot] plot table[x index=0, y index=2] {Data/multTargets/FA_RATE_MIN_NAF_SPACING_NOISE_\noise_ANTENNAS_\antennas_SAMPLING_DOMAIN_DEG_KAPPA_\k.txt};
		\addplot[thin, color=red, solid, mark=square*, mark options={scale=0.75}, mark phase =6, forget plot] plot table[x index=0, y index=3] {Data/multTargets/FA_RATE_MIN_NAF_SPACING_NOISE_\noise_ANTENNAS_\antennas_SAMPLING_DOMAIN_NAF_KAPPA_\k.txt};
		\addplot[thin, color=red, dashed, mark=square*, mark options={scale=0.75}, forget plot] plot table[x index=0, y index=3] {Data/multTargets/FA_RATE_MIN_NAF_SPACING_NOISE_\noise_ANTENNAS_\antennas_SAMPLING_DOMAIN_DEG_KAPPA_\k.txt};



		
\addplot[thin, color=black, dotted, draw=none] coordinates {(0, 1)}; 
\addplot[thin, color=black, solid, draw=none] coordinates {(0, 1)}; 

\end{semilogyaxis} 

\end{tikzpicture}

%% file: Figures/Missed_Det_NAF_Sweep_128.tikz.tex
\if\single1
\def\scale{.635}
\else
\def\scale{.72}
\fi

\tikzset{mark size=2.5}    
    \begin{tikzpicture}
	\begin{semilogyaxis}[
    		xlabel={$\Delta \ell$},
    		ylabel={$P^\text{(MD)}$},
    	ylabel shift = -14 pt,
      	ylabel style={font=\footnotesize,at={(axis description cs:.-0.05,.5)},rotate=0,anchor=south},
      	ytick={0.000001, 0.000002, 0.000003, 0.000004, 0.000005, 0.000006, 0.000007, 0.000008, 0.000009, 0.00001, 0.00002, 0.00003, 0.00004, 0.00005, 0.00006, 0.00007, 0.00008, 0.00009, 0.0001, 0.0002, 0.0003, 0.0004, 0.0005, 0.0006, 0.0007, 0.0008, 0.0009, 0.001, 0.002, 0.003, 0.004, 0.005, 0.006, 0.007, 0.008, 0.009, 0.01, 0.02, 0.03, 0.04, 0.05, 0.06, 0.07, 0.08, 0.09, 0.1, 0.2, 0.3, 0.4},
      	yticklabels={$10^{-6}$, , , , , , , , ,$10^{-5}$, , , , , , , , ,$10^{-4}$, , , , , , , , , , , , , , , , , , $10^{-2}$, , , , , , , , , $10^{-1}$, },
      	xlabel style={font=\footnotesize},
        yticklabel style={font=\footnotesize,xshift=3pt},
        xticklabel style={
        	/pgf/number format/precision=4,
  			/pgf/number format/fixed, 
  			font=\footnotesize},
    	ymin=1e-6,
		ymax=2e-1,
		xmin=0,
		xmax=0.0112,
		minor tick num=1,
		yminorticks = true,
		enlargelimits = false,
		legend pos = south west,
		scaled x ticks = false,
		vasymptote=0.003921,
		grid = both,
    		grid style={solid, black!15},
    		legend columns = 6,
    		transpose legend,
    		legend style={font=\scriptsize},
		xtick = {0, 0.0025, 0.005, 0.0075, 0.01},
		xticklabels = {0, 0.0025, 0.005, 0.0075, 0.01},
        ticklabel style={font=\footnotesize},
		legend cell align={left},
		every axis plot/.append style={very thick},
		mark repeat={7},
		scale = \scale,
		]

\def\antennas{"128"}
\def\kappa{"6.0"}
\def\noise{"-5.0"}
	
		\addplot[thin, color=blue, solid, mark=*, mark options={scale=0.75}, mark phase = 5, forget plot] plot table[x index=0, y index=1] {Data/multTargets/MISSED_DET_MIN_NAF_SPACING_NOISE_\noise_ANTENNAS_\antennas_SAMPLING_DOMAIN_NAF_KAPPA_\kappa.txt};
		\addplot[thin, color=green, solid, mark=*, mark options={scale=0.75}, mark phase = 6, forget plot] plot table[x index=0, y index=2] {Data/multTargets/MISSED_DET_MIN_NAF_SPACING_NOISE_\noise_ANTENNAS_\antennas_SAMPLING_DOMAIN_NAF_KAPPA_\kappa.txt};
		\addplot[thin, color=green, dashed, mark=*, mark options={scale=0.75}, forget plot] plot table[x index=0, y index=2] {Data/multTargets/MISSED_DET_MIN_NAF_SPACING_NOISE_\noise_ANTENNAS_\antennas_SAMPLING_DOMAIN_Deg_KAPPA_\kappa.txt};
		\addplot[thin, color=red, solid, mark=*, mark options={scale=0.75}, forget plot] plot table[x index=0, y index=3] {Data/multTargets/MISSED_DET_MIN_NAF_SPACING_NOISE_\noise_ANTENNAS_\antennas_SAMPLING_DOMAIN_NAF_KAPPA_\kappa.txt};
		\addplot[thin, color=red, dashed, mark=*, mark options={scale=0.75}, forget plot] plot table[x index=0, y index=3] {Data/multTargets/MISSED_DET_MIN_NAF_SPACING_NOISE_\noise_ANTENNAS_\antennas_SAMPLING_DOMAIN_DEG_KAPPA_\kappa.txt};

\def\noise{"-30.0"}

		\addplot[thin, color=blue, solid, mark=square*, mark phase = 4,forget plot] plot table[x index=0, y index=1] {Data/multTargets/MISSED_DET_MIN_NAF_SPACING_NOISE_\noise_ANTENNAS_\antennas_SAMPLING_DOMAIN_NAF_KAPPA_\kappa.txt};
		\addplot[thin, color=green, solid, mark=square*, mark options={scale=0.75}, mark phase = 5, forget plot] plot table[x index=0, y index=2] {Data/multTargets/MISSED_DET_MIN_NAF_SPACING_NOISE_\noise_ANTENNAS_\antennas_SAMPLING_DOMAIN_NAF_KAPPA_\kappa.txt};
		\addplot[thin, color=green, dashed, mark=square*, mark options={scale=0.75}, forget plot] plot table[x index=0, y index=2] {Data/multTargets/MISSED_DET_MIN_NAF_SPACING_NOISE_\noise_ANTENNAS_\antennas_SAMPLING_DOMAIN_Deg_KAPPA_\kappa.txt};
		\addplot[thin, color=red, solid, mark=square*, mark options={scale=0.75}, mark phase = 2, forget plot] plot table[x index=0, y index=3] {Data/multTargets/MISSED_DET_MIN_NAF_SPACING_NOISE_\noise_ANTENNAS_\antennas_SAMPLING_DOMAIN_NAF_KAPPA_\kappa.txt};
		\addplot[thin, color=red, dashed, mark=square*, mark options={scale=0.75}, forget plot] plot table[x index=0, y index=3] {Data/multTargets/MISSED_DET_MIN_NAF_SPACING_NOISE_\noise_ANTENNAS_\antennas_SAMPLING_DOMAIN_DEG_KAPPA_\kappa.txt};


		\addplot[thin, color=blue, solid, draw=none] coordinates {(0, 1)}; \addlegendentry{\gls{SARA} \gls{LAD}}
		\addplot[thin, color=green, solid, draw=none] coordinates {(0, 1)}; \addlegendentry{CUBIC \gls{LAD}}
		\addplot[thin, color=green, dashed, draw=none] coordinates {(0, 1)}; 
		\addplot[thin, color=red, solid, draw=none] coordinates {(0, 1)}; \addlegendentry{MUSIC \gls{LAD}}

		\addplot[thin, color=green, dashed, mark=*, mark options={scale=0.75}, forget plot] plot table[x index=0, y index=2] {Data/multTargets/NAF_RMSE_MIN_NAF_SPACING_\noise_ANTENNAS_\antennas_SAMPLING_DOMAIN_Deg_KAPPA_\kappa.txt};
		
        \addlegendentry{CUBIC Angle}
		\addplot[thin, color=black, solid, mark=*, mark options={scale=0.75}, draw=none] coordinates {(0, 1)};\addlegendentry{$\sigma_n^2 = -5$ dB}
		\addplot[thin, color=black, solid, mark=square*, mark options={scale=0.75}, draw=none] coordinates {(0, 1)};\addlegendentry{$\sigma_n^2 = -30$ dB}

\legend{}   	
\end{semilogyaxis} 

\end{tikzpicture}
	
		

%% file: Figures/RMSE_NAF_Sweep_128.tikz.tex
\if\single1
\def\scale{.635}
\else
\def\scale{.72}
\fi

\tikzset{mark size=2.5}    
    \begin{tikzpicture}
	\begin{semilogyaxis}[
    		xlabel={$\Delta \ell$},
    		ylabel={\gls{LAD} \gls{RMSE}},
    	ylabel shift = -14 pt,
      	ylabel style={font=\footnotesize,at={(axis description cs:.-0.05,.5)},rotate=0,anchor=south},
      	ytick={0.00001, 0.00002, 0.00003, 0.00004, 0.00005, 0.00006, 0.00007, 0.00008, 0.00009, 0.0001, 0.0002, 0.0003, 0.0004, 0.0005, 0.0006, 0.0007, 0.0008, 0.0009, 0.001, 0.002, 0.003, 0.004, 0.005, 0.006, 0.007, 0.008, 0.009, 0.01, 0.02, 0.03, 0.04, 0.05, 0.06, 0.07, 0.08, 0.09, 0.1, 0.2},
      	yticklabels={$10^{-5}$, , , , , , , , , $10^{-4}$, , , , , , , , , , , , , , , , , , , , , , , , , , , $10^{-1}$},
      	xlabel style={font=\footnotesize},
      	yticklabel style={font=\footnotesize,xshift=3pt},
        xticklabel style={
        	/pgf/number format/precision=4,
  			/pgf/number format/fixed, 
  			font=\footnotesize},
    	ymin=1e-5,
		ymax=2e-1,
		xmin=0,
		xmax=0.0112,
  		scaled x ticks = false,
		minor tick num=1,
		yminorticks = true,
		enlargelimits = false,
		legend pos = south west,
		vasymptote=0.003921,
		grid = both,
    		grid style={solid, black!15},
    		legend columns = 6,
    		transpose legend,
    		legend style={font=\scriptsize},
		xtick = {0, 0.0025, 0.005, 0.0075, 0.01},
		xticklabels = {0, 0.0025, 0.005, 0.0075, 0.01},
        ticklabel style={font=\footnotesize},
		legend cell align={left},
		every axis plot/.append style={very thick},
		mark repeat={7},
		scale = \scale
		]

\def\antennas{"128"}
\def\kappa{"6.0"}
\def\noise{"-5.0"}
	
		\addplot[thin, color=blue, solid, mark=*, mark options={scale=0.75},  forget plot] plot table[x index=0, y index=1] {Data/multTargets/NAF_RMSE_MIN_NAF_SPACING_\noise_ANTENNAS_\antennas_SAMPLING_DOMAIN_NAF_KAPPA_\kappa.txt};
		\addplot[thin, color=green, solid, mark=*, mark options={scale=0.75}, forget plot] plot table[x index=0, y index=2] {Data/multTargets/NAF_RMSE_MIN_NAF_SPACING_\noise_ANTENNAS_\antennas_SAMPLING_DOMAIN_NAF_KAPPA_\kappa.txt};
		\addplot[thin, color=green, dashed, mark=*, mark options={scale=0.75}, forget plot] plot table[x index=0, y index=2] {Data/multTargets/NAF_RMSE_MIN_NAF_SPACING_\noise_ANTENNAS_\antennas_SAMPLING_DOMAIN_Deg_KAPPA_\kappa.txt};
		\addplot[thin, color=red, solid, mark=*, mark options={scale=0.75}, forget plot] plot table[x index=0, y index=3] {Data/multTargets/NAF_RMSE_MIN_NAF_SPACING_\noise_ANTENNAS_\antennas_SAMPLING_DOMAIN_NAF_KAPPA_\kappa.txt};
		\addplot[thin, color=red, dashed, mark=*, mark options={scale=0.75}, forget plot] plot table[x index=0, y index=3] {Data/multTargets/NAF_RMSE_MIN_NAF_SPACING_\noise_ANTENNAS_\antennas_SAMPLING_DOMAIN_DEG_KAPPA_\kappa.txt};

\def\noise{"-30.0"}

		\addplot[thin, color=blue, solid, mark=square*, mark options={scale=0.75}, forget plot] plot table[x index=0, y index=1] {Data/multTargets/NAF_RMSE_MIN_NAF_SPACING_\noise_ANTENNAS_\antennas_SAMPLING_DOMAIN_NAF_KAPPA_\kappa.txt};
		\addplot[thin, color=green, solid, mark=square*, mark options={scale=0.75}, forget plot] plot table[x index=0, y index=2] {Data/multTargets/NAF_RMSE_MIN_NAF_SPACING_\noise_ANTENNAS_\antennas_SAMPLING_DOMAIN_NAF_KAPPA_\kappa.txt};
		\addplot[thin, color=green, dashed, mark=square*, mark options={scale=0.75}, forget plot] plot table[x index=0, y index=2] {Data/multTargets/NAF_RMSE_MIN_NAF_SPACING_\noise_ANTENNAS_\antennas_SAMPLING_DOMAIN_Deg_KAPPA_\kappa.txt};
		\addplot[thin, color=red, solid, mark=square*, mark options={scale=0.75}, forget plot] plot table[x index=0, y index=3] {Data/multTargets/NAF_RMSE_MIN_NAF_SPACING_\noise_ANTENNAS_\antennas_SAMPLING_DOMAIN_NAF_KAPPA_\kappa.txt};
		\addplot[thin, color=red, dashed, mark=square*, mark options={scale=0.75}, forget plot] plot table[x index=0, y index=3] {Data/multTargets/NAF_RMSE_MIN_NAF_SPACING_\noise_ANTENNAS_\antennas_SAMPLING_DOMAIN_DEG_KAPPA_\kappa.txt};

\legend{}

\end{semilogyaxis} 

\end{tikzpicture}
	

%% file: Content/2DExample.tex
\subsection{\gls{2D} imaging example}

\begin{figure*}[h!]
        \centering
        \begin{subfigure}{0.32\textwidth}
            \centering
			\includegraphics[width=1\textwidth]{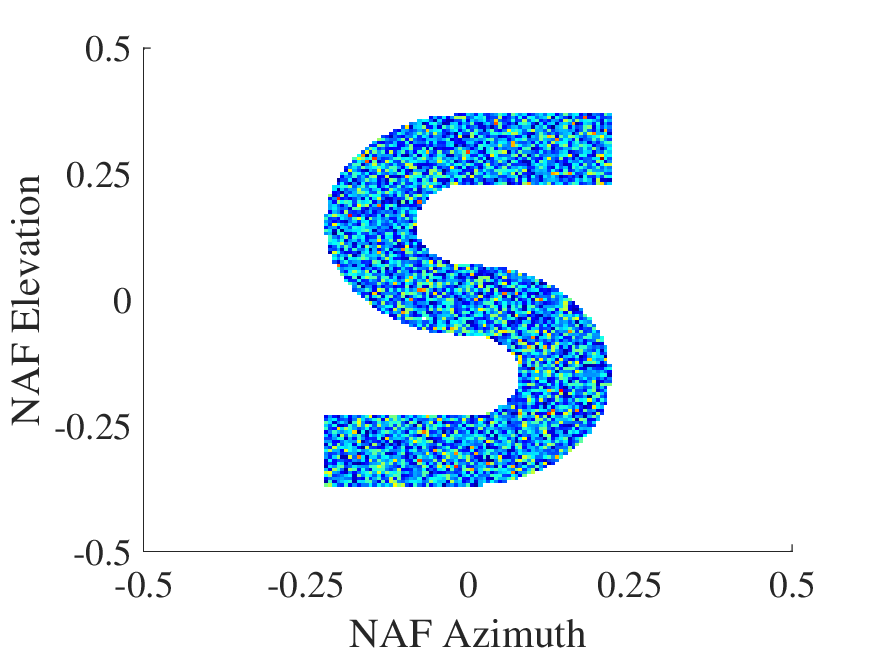}
            \caption{Real Scenario with $5017$ point scatterers.} 
            \label{subfig:2DImage_Ideal}
        \end{subfigure}
        \hfill
        \begin{subfigure}{0.32\textwidth}
            \centering
			\includegraphics[width=1\textwidth]{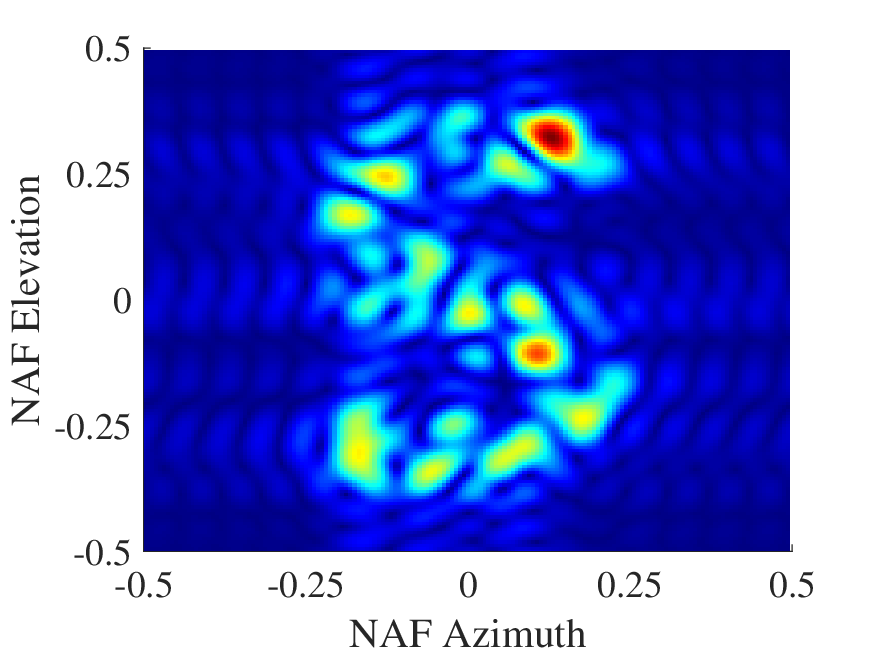}
            \caption{Response sampled at 160x160 \gls{2D} angles.} 
            \label{subfig:2DImage_ArrayIdeal}
        \end{subfigure}
        \hfill
        \begin{subfigure}{0.32\textwidth}
            \centering
			\includegraphics[width=1\textwidth]{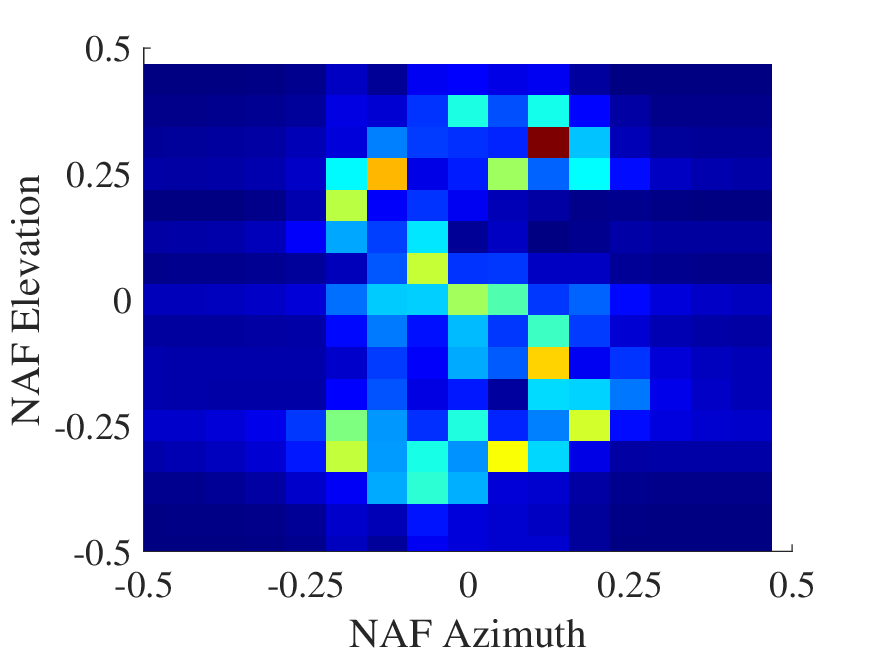}
            \caption{Response sampled at 16x16 \gls{2D} angles.} 
            \label{subfig:2DImage_Sampled}
        \end{subfigure}
        \vskip\baselineskip
        \begin{subfigure}{0.32\textwidth}
            \centering
			\includegraphics[width=1\textwidth]{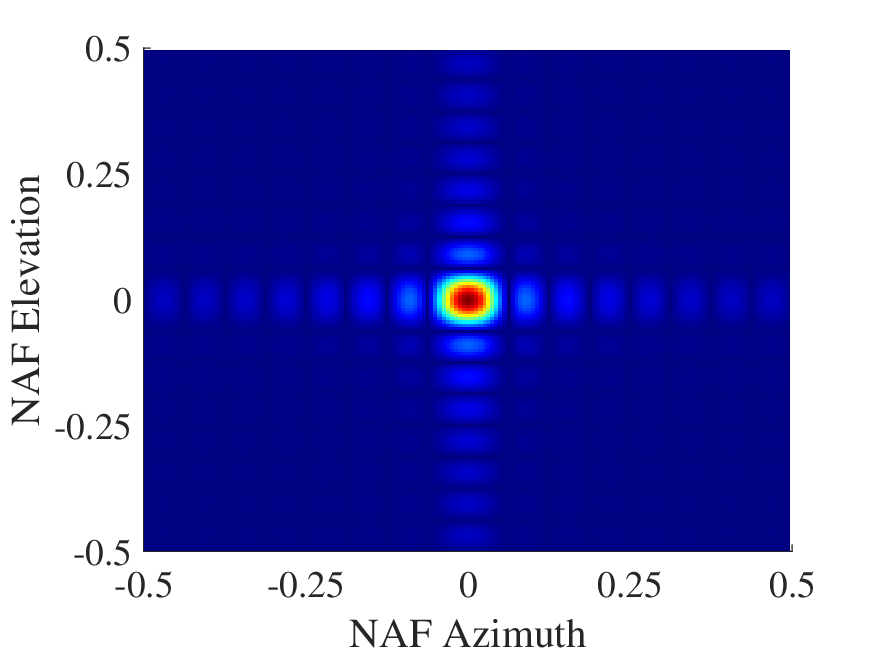}
            \caption{Dirichlet kernel used for reconstruction.} 
            \label{subfig:2DImage_Dirichlet}
        \end{subfigure}
        \hfill
        \begin{subfigure}{0.32\textwidth}
            \centering
			\includegraphics[width=1\textwidth]{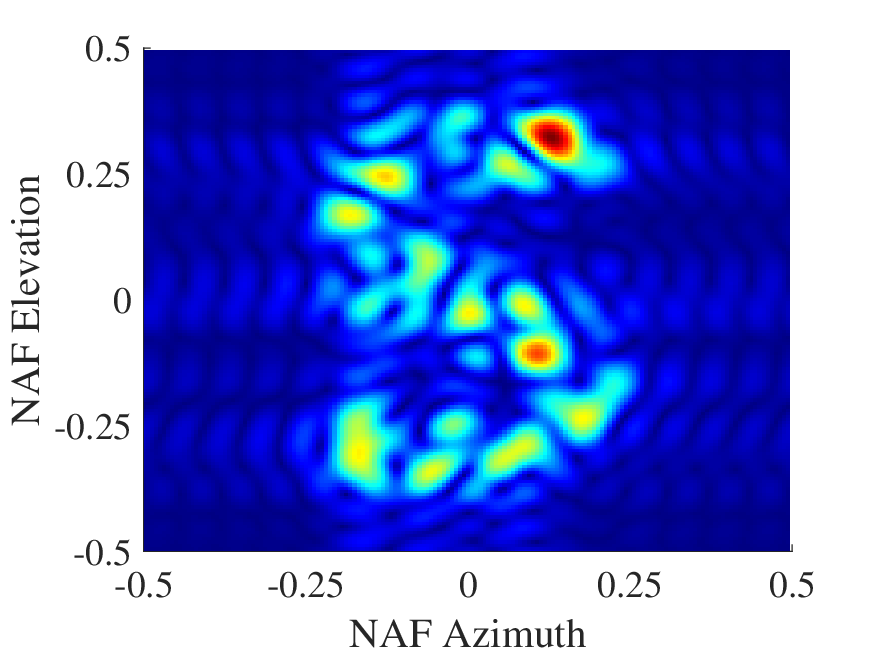}
            \caption{The upsampled response with \gls{SARA}.} 
            \label{subfig:2DImage_Reconstructed_SARA}
        \end{subfigure}
        \hfill
        \begin{subfigure}{0.32\textwidth}
            \centering
			\includegraphics[width=1\textwidth]{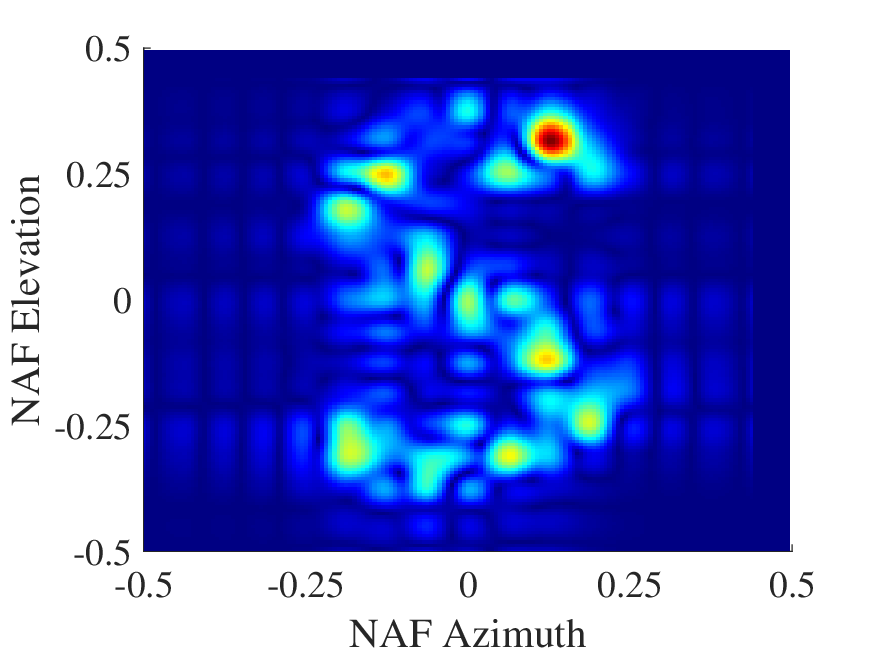}
            \caption{The upsampled response with CUBIC.} 
            \label{subfig:2DImage_Reconstructed_CUBIC}
        \end{subfigure}

        \caption{Example of imaging with a receiver consisting of a 16x16 elements \gls{URA} and $d = \lambda/2$, with omni-directional scenario illumination. All figures plot the magnitude of the respective quantity in jet scale (blue lower values, red higher values). The real scenario is distorted by the limited aperture of the array, that is sampled with a fine (160x160 \gls{2D} angles) and a coarse (16x16 \gls{2D} angles) grid. Then, a Dirichlet kernel is used to interpolate the coarse scan, upsampling it to 160x160 with \gls{SARA}. For comparison, also the result of cubic interpolation is shown. 
} 
\label{fig:2D_Image}
\end{figure*}

We conclude the numerical evaluations in this paper by providing a \gls{2D} imaging example in Fig.~\ref{fig:2D_Image}. Along the lines of~\cite{rajamaki2020hybrid} and similar works, we consider the worst-case scenario of an object with $Q$ \gls{iid} point scatterers, each following a complex normal distribution: $\mathcal{CN}\left( \frac{1}{\sqrt{2Q}}, \frac{1}{\sqrt{2Q}}\right)$. In our considered example, the true shape can be found in Fig.~\ref{subfig:2DImage_Ideal}, where $Q=5017$ and we plot axes in the \gls{LAD} domain.

Assuming that no noise is present, we assume to illuminate the area with an omni-directional signal, allowing us to focus on the receiver's impact only. A 16x16 \gls{URA} with antenna spacing $d = \lambda/2$ is considered. The limited aperture of the array generates a loss of resolution, as illustrated in Section \ref{sec:ArrayAngularDomains}. This is visible in Fig.~\ref{subfig:2DImage_ArrayIdeal}, where the absolute value of the response of the array is plotted, scanning the 
azimuth and elevation \gls{LAD} pairs - or \gls{2D} angle - $\boldsymbol{\eta}$ in the set $\mathcal{L}_{160,160}^{\text{(Rx)}}$ defined in \eqref{eq:URA_Angles}. This means that a total of 25600 samples is required. In practical wireless systems, this is undesired for the great consumption of wireless resources needed to perform so many scans.
Moreover, if we considered the scenario of Table~\ref{tab:SimDefaultParameters}, this would require a full scan acquisition time in the order of hundreds of milliseconds, that would generate phase distortions in dynamic scenarios, impacting the accuracy of the reconstruction, as seen in \ref{subsec:SimSingleTarget}, paragraph c).

Therefore, we sample the response at 16x16 \gls{2D} angles $\boldsymbol{\eta} \in \mathcal{L}_{16,16}^{\text{(Rx)}}$, producing the low-resolution sampled response shown in Fig.~\ref{subfig:2DImage_Sampled}. Then, applying what was discussed in Subsection \ref{subsec:URAExtension}, we can derive the Dirichlet kernel, plotted in Fig.~\ref{subfig:2DImage_Dirichlet}, and use it to upsample the \gls{2D} image by a factor 10. Comparing the initial 160x160 scan of Fig.~\ref{subfig:2DImage_ArrayIdeal} and the reconstructed response in Fig.~\ref{subfig:2DImage_Reconstructed_SARA}, one can notice that they are exactly the same. For comparison, also the reconstruction obtained by cubic interpolation is shown in Fig.~\ref{subfig:2DImage_Reconstructed_CUBIC}, where the distortion introduced is clearly visible. In particular, defining the normalized \gls{RMSE} of the reconstruction as
\begin{equation}
\epsilon = \sqrt{\frac{\sum_{\boldsymbol{\eta} \in \mathcal{L}_{160,160}^{\text{(Rx)}}} \left[ \left | R(\boldsymbol{\eta}) - L(\boldsymbol{\eta}) \right |^2 \right]}
{\sum_{\boldsymbol{\eta} \in \mathcal{L}_{160,160}^{\text{(Rx)}}} \left[ \left | L(\boldsymbol{\eta}) \right |^2 \right]}} \;,
\end{equation}
we obtained $\epsilon^\text{(\gls{SARA})} = 0$ and $\epsilon^\text{(CUBIC)} \approx 40\%$ with \gls{SARA} and cubic reconstruction, respectively.

%% file: Content/Conclusion.tex
\section{Conclusion}\label{sec:Conclusion}
This paper provided the necessary theory to perfectly sample and reconstruct the angular response (SARA) of \gls{ULA} and \gls{URA}. This is a problem of particular relevance when operating with analog and hybrid arrays, that requires to focus on specific angle(s) at each signal transmission/reception. 
The minimum number and the set of angles to be scanned have been derived, and the Whittaker-Shannon interpolation theorem has been leveraged to demonstrate that the perfect reconstruction of the full angular response of the array coincides with a proper trigonometric interpolation in the normalized angular frequency.
Extensions to cases with more or fewer scans, as well as practical applications of theory to sensing use cases are provided, elaborating on the particular case of a full duplex transmit and receive array system.

The performance of the \gls{SARA} proposal has been evaluated against prior art angular sampling and reconstruction techniques, on angular estimation tasks of single and multiple targets. \gls{SARA} consistently outperforms the reconstruction baseline solutions investigated, achieving similar performance as the super-resolution MUSIC algorithm for single impulsive target angular estimation, with the additional advantage of important savings in computational complexity. Moreover, our proposed \gls{DFT}-based interpolation lowers the required complexity of interpolation below the sub-optimal cubic spline baseline, without any distortion in the reconstruction.
We have shown that \gls{SARA} even approaches the Cram\'er-Rao lower bound approximation of angular estimation achievable with fully digital beamforming at high \gls{SNR}.

When considering multiple targets, \gls{SARA}  improves the \gls{LAD} \gls{RMSE} at high target separation compared to reconstruction and super-resolution baselines, especially at high \gls{SNR}. For closer targets, MUSIC offers better resolution capabilities due to is super-resolution properties, but experiences higher false alarm rates and exhibits a severe performance degradation in mid to high noise scenarios. Compared to the cubic reconstruction algorithm, \gls{SARA} yields gains with regard to multi-target resolution and \gls{LAD} \gls{RMSE}, while keeping false alarms at an arbitrary desired level. 


In our \gls{2D} imaging example, \gls{SARA}'s perfect sampling and reconstruction can avoid the distortion induced by other baselines, which experiences a normalized \gls{RMSE} of $40\%$.

In future work, we plan to leverage \gls{SARA} to showcase its imaging capabilities in an \gls{JCAS} proof of concept, that extends current \gls{5G} hardware deployments by incorporating radar capabilities. Moreover, we plan to extend the considerations done in this work for narrowband signals to a broadband signal model.

%% file: Content/Acknowledgements.tex
\section*{Acknowledgments}
The authors want to thank Stefan Wesemann, Thorsten Wild and Traian Emanuel Abrudan for their helpful suggestions, insights, and technical discussions.

%% file: Content/Appendixes.tex
\begin{appendices}
\appendices

\section{Proof of Lemma \ref{lemma:PlanarWaveResponse}}\label{app:PlanarWaveResponse}
The \gls{LAD} response of an incident planar wave with \gls{LAD} $\eta = 0$ is
\begin{align}
L_N(\ell, \eta) &= \frac{1}{N}\sum_{n = 0}^{N-1} a_n(0) e^{-j {2\pi}\frac{x_n}{d} \ell}
= \frac{1}{N}\sum_{n = 0}^{N-1} e^{-j {2\pi} \frac{x_n}{d} \ell} 
\; .
\label{eq:DirichletEffectZero}
\end{align}
One can substitute \eqref{eq:ArrayAntennasLocations} into \eqref{eq:DirichletEffectZero}, having that
\begin{equation}
L(\ell, 0) =
\frac{1}{N}e^{j 2 \pi \frac{N-1}{2} \ell}
\sum_{n=0}^{N-1} e^{-j 2 \pi \ell n} \; .
\label{app_eq:LAD1}
\end{equation}
Leveraging the known formula on truncated geometric series, we have
\if\single1
\begin{align}
b \sum_{n=0}^{N-1} r^n = b \frac{1 - r^N}{1 - r} = 
b \frac{r^{\frac{N}{2}}}{r^{\frac{1}{2}}}  \cdot
\frac{r^{-\frac{N}{2}} - r^{\frac{N}{2}}}{r^{-\frac{1}{2}} - r^{-\frac{1}{2}}}
= b r^{\frac{N-1}{2}}
\frac{r^{-\frac{N}{2}} - r^{\frac{N}{2}}}{r^{-\frac{1}{2}} - r^{-\frac{1}{2}}}
 \; .
\label{app_eq:LAD2}
\end{align}
In particular, \eqref{app_eq:LAD2} is equivalent to \eqref{app_eq:LAD1} 
\else
\begin{align}
b \sum_{n=0}^{N-1} r^n &= b \frac{1 - r^N}{1 - r} = 
b \frac{r^{\frac{N}{2}}}{r^{\frac{1}{2}}}  \cdot
\frac{r^{-\frac{N}{2}} - r^{\frac{N}{2}}}{r^{-\frac{1}{2}} - r^{-\frac{1}{2}}}
= \nonumber \\
&= b r^{\frac{N-1}{2}}
\frac{r^{-\frac{N}{2}} - r^{\frac{N}{2}}}{r^{-\frac{1}{2}} - r^{-\frac{1}{2}}}
 \; .
\label{app_eq:LAD2}
\end{align}
Note that \eqref{app_eq:LAD2} is equivalent to \eqref{app_eq:LAD1} 
\fi
with $b = N^{-1} e^{j 2 \pi  \frac{N-1}{2} \ell}$ and $r = e^{-j 2 \pi  \ell}$.
Note that $br^{\frac{N-1}{2}} = N^{-1}$, thus
\if\single1
\begin{align}
L(\ell, 0) &= 
\frac{1}{N}
\frac{
e^{+j 2 \pi  \ell \frac{N}{2}} - 
e^{-j 2 \pi  \ell \frac{N}{2}}
}{
e^{+j 2 \pi  \ell \frac{1}{2}} - 
e^{-j 2 \pi  \ell \frac{1}{2}}
}= 
\frac{
\sin \left( 2 \pi \ell \frac{N}{2}\right)
}{N
\sin \left( 2 \pi \ell \frac{1}{2}\right)
} = 
\frac{
\sin \left( \pi N \ell \right)
}{N
\sin \left( \pi \ell \right)
} \;.
\label{app_eq:LAD3}
\end{align}
\else
\begin{align}
L(\ell, 0) &= 
\frac{1}{N}
\frac{
e^{+j 2 \pi  \ell \frac{N}{2}} - 
e^{-j 2 \pi  \ell \frac{N}{2}}
}{
e^{+j 2 \pi  \ell \frac{1}{2}} - 
e^{-j 2 \pi  \ell \frac{1}{2}}
}= 
\frac{
\sin \left( 2 \pi \ell \frac{N}{2}\right)
}{N
\sin \left( 2 \pi \ell \frac{1}{2}\right)
} = \nonumber \\
&=
\frac{
\sin \left( \pi N \ell \right)
}{N
\sin \left( \pi \ell \right)
} \;.
\label{app_eq:LAD3}
\end{align}
\fi
We now consider a generic \gls{LAD} $\eta$
\begin{align}
L_N(\ell, \eta) &= \frac{1}{N}\sum_{n = 1}^{N} a_n(\eta) e^{-j {2\pi}\frac{x_n}{d} \ell} = \frac{1}{N}\sum_{n = 0}^{N-1} e^{-j {2\pi} \frac{x_n}{d} (\ell-\eta)}\;.
\label{eq:lemmaPlanarWaveResponse2}
\end{align}
One can immediately notice that \eqref{eq:lemmaPlanarWaveResponse2} is the translation of \eqref{eq:DirichletEffectZero} by a term $\eta$, therefore
\begin{equation}
L_N(\ell, \eta) = \frac{\sin \left( \pi N (\ell - \eta)  \right)}{N \sin \left( \pi (\ell - \eta) \right)} \; .
\end{equation}


\section{Proof of Lemma \ref{lemma:NAFPeriodicity}}\label{app:NAFPeriodicity}
One can use \eqref{eq:RxBeamformerShifted} to expand the array response as follows
\if\single1
\begin{align}
L'_N\left(\ell - k \right) &=
\frac{1}{N} \sum_{n=0}^{N-1} a_n e^{-j 2\pi\left(x'_n + \frac{N-1}{2}\right) \left(\ell - k \right)} 
=
\frac{1}{N} \sum_{n=0}^{N-1} a_n e^{-j 2\pi\left(x'_n +  \frac{N-1}{2}\right)\ell}
e^{j2\pi (-\frac{N-1}{2} + n + \frac{N-1}{2})k } = \nonumber \\
&= 
\frac{1}{N} \sum_{n=0}^{N-1} a_n e^{-j 2\pi\left(x'_n +  \frac{N-1}{2}\right)\ell}e^{j2\pi nk} 
\overset{k \in \mathbb{Z}}{=} \frac{1}{N} \sum_{n=0}^{N-1} a_n e^{-j 2\pi\left(x'_n +  \frac{N-1}{2}\right)\ell} = L'_N(\ell) \; .
\end{align}
\else
\begin{align}
L'_N\left(\ell - k \right) &=
\frac{1}{N} \sum_{n=0}^{N-1} a_n e^{-j 2\pi\left(x'_n + \frac{N-1}{2}\right) \left(\ell - k \right)}
= \nonumber \\
&= 
\frac{1}{N} \sum_{n=0}^{N-1} a_n e^{-j 2\pi\left(x'_n +  \frac{N-1}{2}\right)\ell}\cdot
\nonumber \\
&\cdot
e^{j2\pi (-\frac{N-1}{2} + n + \frac{N-1}{2})k } = \nonumber \\
&= 
\frac{1}{N} \sum_{n=0}^{N-1} a_n e^{-j 2\pi\left(x'_n +  \frac{N-1}{2}\right)\ell}e^{j2\pi nk} = \nonumber \\
&\overset{k \in \mathbb{Z}}{=} \frac{1}{N} \sum_{n=0}^{N-1} a_n e^{-j 2\pi\left(x'_n +  \frac{N-1}{2}\right)\ell} = L'_N(\ell) \; .
\end{align}
\fi


\section{Infinite summation of \eqref{eq:InterpolationSincStep1}}\label{app:InfiniteSinc}
In this appendix section, we want to simplify the following
\if\single1
\begin{align}
s_N(\ell) &= \sum_{k = -\infty}^{+\infty} e^{-j2\pi \frac{N-1}{2}  \left(\ell - k  \right)}
\text{sinc} \left( N \left(\ell - k \right) \right) =
e^{-j2\pi \frac{N-1}{2}  \ell}\text{sinc}(N\ell) * \sum_{k = -\infty}^{+\infty} 
\delta \left(\ell - k \right) 
\;,
\label{app_eq:Sinc1}
\end{align}
\else
 summation
\begin{align}
s_N(\ell) &= \sum_{k = -\infty}^{+\infty} e^{-j2\pi \frac{N-1}{2}  \left(\ell - k  \right)}
\text{sinc} \left( N \left(\ell - k \right) \right) =
\nonumber \\
&=
e^{-j2\pi \frac{N-1}{2}  \ell}\text{sinc}(N\ell) * \sum_{k = -\infty}^{+\infty} 
\delta \left(\ell - k \right) 
\;,
\label{app_eq:Sinc1}
\end{align}
\fi
where $*$ and $\delta(\ell)$ are the linear convolution and the Dirac impulse function, respectively. Due to the convolution becoming a multiplication in the Fourier domain and vice versa, if one considers the inverse Fourier transform of \eqref{app_eq:Sinc1}, one has
\if\single1
\begin{align}
S_N(x) &= 
\left( \delta \left(x - \frac{(N-1)}{2} \right) *
\frac{1}{N} \text{rect} \left( \frac{x}{N} \right) \right)  
\frac{1}{2}\sum_{n = \infty}^{+\infty} \delta \left(x - n \right) = \nonumber \\
&= \text{rect} \left( \frac{x}{N} - \frac{N-1}{2N} \right)  
\cdot   
\frac{1}{N}\sum_{n = \infty}^{+\infty} \delta \left(x - n 
\right) \; .
\label{app_eq:SincTransform1}
\end{align}
\else
\begin{align}
S_N(x) &= 
\left( \delta \left(x - \frac{(N-1)}{2} \right) *
\frac{1}{N} \text{rect} \left( \frac{x}{N} \right) \right)  \cdot
\nonumber \\
&\cdot
\frac{1}{2}\sum_{n = -\infty}^{+\infty} \delta \left(x - n \right) = \nonumber \\
&= \text{rect} \left( \frac{x}{N} - \frac{N-1}{2N} \right)  
\cdot   
\frac{1}{N}\sum_{n = -\infty}^{+\infty} \delta \left(x - n 
\right) \;
.
\label{app_eq:SincTransform1}
\end{align}
\fi
The multiplication with the rectangle function has the effect of limiting the series in \eqref{app_eq:SincTransform1} into
\begin{align}
S_N(x) &= 
\frac{1}{N}\sum_{n = 0}^{N-1} \delta \left(x - n \right)
\; .
\label{app_eq:SincTransform2}
\end{align}
Leveraging with some modifications the same considerations of Appendix \ref{app:NAFPeriodicity}, one can apply the Fourier transform back from~\eqref{app_eq:SincTransform2}, obtaining the Dirichlet kernel shape with the linear phase component due to the \gls{AAL} translation

\if\single1
\begin{align}
s_N(\ell) &= 
\frac{1}{N}\sum_{n = 0}^{N-1}
e^{-j2\pi n\ell}
=
e^{-j2\pi \frac{N-1}{2}\ell}
\left(e^{j2\pi \frac{N-1}{2}\ell}
\frac{1}{N}\sum_{n = 0}^{N-1}
e^{-j2\pi 2n\ell} \right)
=
e^{-j2\pi \frac{N-1}{2}\ell}
 D_N \left(\ell \right)  = D'_N \left(\ell \right) \;.
 \label{eq:SincAuxiliaryFinal}
\end{align}
\else
\begin{align}
s_N(\ell) &= 
\frac{1}{N}\sum_{n = 0}^{N-1}
e^{-j2\pi n\ell}
=
\nonumber \\
&=
e^{-j2\pi \frac{N-1}{2}\ell}
\left(e^{j2\pi \frac{N-1}{2}\ell}
\frac{1}{N}\sum_{n = 0}^{N-1}
e^{-j2\pi 2n\ell} \right)
=
\nonumber \\
&=
e^{-j2\pi \frac{N-1}{2}\ell}
 D_N \left(\ell \right)  = D'_N \left(\ell \right) \;.
 \label{eq:SincAuxiliaryFinal}
\end{align}
\fi

\section{IDFT derivations}\label{app:IdftDerivation}
The response IDFT in~\eqref{eq:ResponseDFT} can be obtained by
\begin{align}
\mathbf{l'}_{\mathbf{A},k} &= \sum_{u=0}^{NU-1}\mathbf{l}'_u e^{j 2 \pi \frac{ku}{NU}} = 
\sum_{n=0}^{N-1} \mathbf{\overline{l}'}_n e^{j 2 \pi \frac{kn}{N}} \;, k = 0, ..., NU-1 \;,
\end{align}
which is the $p$-th element of the IDFT with $N$ elements of $\left(\mathbf{\overline{l}'} \right)$, with $p = \text{mod}_N(k)$. Therefore, \eqref{eq:ResponseDFT} is the sequential repetition of the $N$ elements of $ \text{IDFT}\left(\mathbf{\overline{l}'} \right)$, exactly $U$ times.
Regarding the kernel, from the IDFT definition and~\eqref{eq:SincAuxiliaryFinal}, one has
\if\single1
\begin{align}
\mathbf{d'}_{\mathbf{A},k} &= \sum_{u=0}^{NU-1} D'_N \left( \frac{u}{NU} \right)  e^{j 2 \pi \frac{ku}{NU}}
= 
\sum_{u=0}^{NU-1} \sum_{n=0}^{N-1} e^{-j 2 \pi n \frac{u}{NU}}  e^{j 2 \pi \frac{ku}{NU}}
= 
\sum_{u=0}^{NU-1} \sum_{n=0}^{N-1}   e^{j 2 \pi \frac{u(k-n)}{NU}}
=
\nonumber \\
&=
\begin{cases}
1 & \text{ if } 0 \leq k < N \\
0 & \text{ if } N \leq k < NU-1 
\end{cases} \;.
\end{align}
\else
\begin{align}
\mathbf{d'}_{\mathbf{A},k} &= \sum_{u=0}^{NU-1} D'_N \left( \frac{u}{NU} \right)  e^{j 2 \pi \frac{ku}{NU}}
= \nonumber \\
&= \sum_{u=0}^{NU-1} \sum_{n=0}^{N-1} e^{-j 2 \pi n \frac{u}{NU}}  e^{j 2 \pi \frac{ku}{NU}}
= \nonumber \\
&= \sum_{u=0}^{NU-1} \sum_{n=0}^{N-1}   e^{j 2 \pi \frac{u(k-n)}{NU}}
= \nonumber \\
&=
\begin{cases}
1 & \text{ if } 0 \leq k < N \\
0 & \text{ if } N \leq k < NU-1 
\end{cases} \;.
\end{align}
\fi

\section{Derivation of \gls{CFAR} threshold}\label{app:CFARDerivation}

Recalling \eqref{eq:NoisePower}, the noise power is $\sigma^2 = N^{-1}\sigma_n^2$ and distributed as a complex Gaussian variable.
Therefore, the probability that a single noise sample exceeds a threshold $\zeta$ - corresponding to the desired false alarm probability - is given by the \gls{CCDF} of the Rayleigh distribution
\begin{equation}
P^\text{(FA)}_{s} = e^{-\frac{\zeta^2}{\sigma^2}} \;.
\label{app_eq:P_FA_sample}
\end{equation}
Solving (\ref{app_eq:P_FA_sample}) for $\zeta$ yields 
\begin{equation}
\zeta = \sqrt{-\sigma^2\ln(P^\text{(FA)}_{s})} \;.
\label{app_eq:zeta_sample}
\end{equation}
Therefore, the false alarm probability for an angular scan of $2N-1$ samples is
\begin{equation}
P^\text{(FA)} = 1-(1-P^\text{(FA)}_{s})^{2N-1} \;.
\label{app_eq:P_FA}
\end{equation}
Solving for $P^\text{(FA)}_{s}$ and substituting into (\ref{app_eq:zeta_sample}), the \gls{CFAR} threshold for a desired $P^\text{(FA)}$ is
\begin{equation}
\zeta= \sqrt{-\sigma^2\ln(1-(P^\text{(FA)} - 1)^{\frac{1}{2N-1}})} \;.
\label{app_eq:zeta}
\end{equation}
\end{appendices}

%% file: Content/RevisionAnswers_secondReview.tex
\clearpage

\section*{Answer to the Editor and Reviewers' Comments}

The authors thank the Editor and the Reviewers for their comments and support which help to improve the paper. 
We have addressed each of the remaining comments raised and provided a point-by-point response below.   Substantial text changes  are highlighted as \underline{underlined texts}
 in the main body and are subsequently quoted in our responses below.  

\section*{Editor}

~\\ \noindent{\bf (1) There are some further comments by the reviewers based on the revision and response, which require more clear discussion or check. These require some minor revisions to the current manuscript. Please address these comments carefully in the revision. 
 }
\setstretch{1.5} 

~\\ \noindent{\bf Ans:} Thanks for your comment. We tried to address the remaining concerns focusing on the points below.
\begin{itemize}
    \item We further improved the Introduction to further clarify the motivation and its novelty w.r.t. [13] and [17], as requested by Reviewer 3.
    \item We reviewed the reference style, and added missing information. We believe it is now in good shape, as judged by its conformance with the references of the 5 top popular papers listed in TWC webpage.
    \item We reshaped the results section, removing the TSDCE baseline, agreeing with the comment of Reviewer 2.
\end{itemize}

\section*{Reviewer: 1}

~

\setstretch{1.0}

\noindent {\bf The authors have addressed all of comments from the reviewer. The reviewer does not the further suggestions.}
\setstretch{1.0}

~\\ \noindent{\bf  Ans:} We are happy that we have been able to address all of the reviewer's comments, which helped improving the paper.

\newpage
\section*{Reviewer: 2}

~

\setstretch{1.0}

\noindent {\bf First, the reviewer would like to thank the authors for the effort in implementing the reviewer's suggestions.

Regarding the implementation of the TSDCE, and following the discussion presented in the current manuscript and response cover letter, it is clear that this algorithm cannot be properly implemented for comparison in this use case.  

The reviewer appreciates the time and effort that the authors used to try to achieve a fair implementation, but the performance with respect to Cramer Bound does not justifies the noise term penalty, since there are other implementation parameters that might be affecting the performance. 

As the authors pointed out, TSDCE is optimized for a different scenario/use case, so the reviewer agrees that it can be placed in the state-of-the-art discussion, but it should be removed from the simulation results. 
}
\setstretch{1.0}

~\\ \noindent{\bf  Ans:} Agree. We have removed {TSDCE} from the simulations section and added the following text in the Introduction about \cite{roger2021fast,roger2021low}, justifying why TSDCE is not further considered in the paper:

``\revp{rev:TSDCE}''
\setstretch{1.5}

We thank the reviewer for his/her previous comments, that helped us shedding more light and context to our work, improving our understanding and its robustness.

\newpage
\section*{Reviewer: 3}

~

\setstretch{1.0}

\noindent {\bf The authors have largely clarified the questions proposed by the reviewer. Nevertheless, there are a few more points that need to be considered:}
\setstretch{1.0}

~\\ \noindent{\bf 1) The authors have revised Section I, however, the motivation of this paper is still not clear enough, and more improvements should be provided.  }
\setstretch{1.5}

~\\ \noindent{\bf  Ans:} We thank the reviewer for the feedback. To further clarify the motivation, we added a ``Motivation'' Subsection in the Introduction by compiling paragraphs, reorganizing and rephrasing it, to highlight the communication orientated nature of existing beam scanning approaches in the literature and to reflect the need of imaging that 6G sensing requires.  We further discussed the shortcomings of radio imaging methods considered in the analog/hybrid beamforming literature, which typically rely on the sparsity of the (i) channel estimation or (ii) direction of departure/arrival estimation. 
We hope that the changes in this new Subsection~\ref{subsec:Motivation} go in the right direction, offering a better context for the problem we are solving, that is described in the following text

``\revp{rev:ProblemSummary}''

\setstretch{1.5}

~\\ \noindent{\bf 2) Although the authors have added [13] and [17] as new references, the clarity of novelty is worth exploring.  }
\setstretch{1.5}

~\\ \noindent{\bf  Ans:} Thanks for the suggestion. We have expanded text, hopefully clarifying the difference between prior art's contribution and our work. Note that, after the review, the new numbering of the references is~\cite{lin2020nested,lin2021joint}. Accordingly, the paragraph in the introduction has been rephrased as follows

``\revp{rev:newRef2}''

\setstretch{1.5}

~\\ \noindent{\bf 3) The reference style does not seem to have changed much, and authors should refer to papers already published in TWC for revision. }
\setstretch{1.5}

~\\ \noindent{\bf  Ans:} We thank the reviewer for the comment. In particular, we noticed that the month was missing from our references and that the we referenced books in a wrong way. We also revised and equalized the reference style of conference proceedings. Moreover, some other cosmetic changes were done to the reference section. After checking with recent TWC publications, the reference section now seems to be in line with the TWC reference style.

\setstretch{1.5}

~\\ \noindent{\bf 4) The legends in Fig. 8 and Fig. 9 are confusing. The legends in Fig. 8 should be consistent with the legends in Fig.9. }
\setstretch{1.5}

~\\ \noindent{\bf  Ans:} We thank the reviewer for spotting this inconsistency. We have harmonized the legend style (colors for algorithm, dash style for sampling criterion, marker for noise power) in Figs. 8 and 9. 
\setstretch{1.5}